\documentclass[twocolumn]{aastex631} 

\newcommand{\kms}{km\,s$^{-1}$}

\newcommand{\discminer}{\textsc{discminer}}
\newcommand{\disksurf}{\textsc{disksurf}}

\newcommand{\exoalma}{exoALMA}

\newcommand{\twCOfull}{$^{12}$CO\,$J=3-2$}
\newcommand{\twCO}{$^{12}$CO}
\newcommand{\thCO}{$^{13}$CO}
\newcommand{\thCOfull}{$^{13}$CO\,$J=3-2$}

\newcommand{\CSfull}{CS\,$J=7-6$}

\newcommand{\aatau}{AA\,Tau}
\newcommand{\cqtau}{CQ\,Tau}
\newcommand{\dmtau}{DM\,Tau}
\newcommand{\hdthree}{HD\,34282}
\newcommand{\hdone}{HD\,135344B}
\newcommand{\hdonefour}{HD\,143006}
\newcommand{\jfour}{J1604}
\newcommand{\jfifteen}{J1615}
\newcommand{\jfortytwo}{J1842}
\newcommand{\jfiftytwo}{J1852}
\newcommand{\lkca}{LkCa\,15}
\newcommand{\mwcsev}{MWC\,758}
\newcommand{\pdssix}{PDS\,66}
\newcommand{\sycha}{SY\,Cha}
\newcommand{\vforty}{V4046\,Sgr}

\usepackage[]{hyperref}
\hypersetup{
  colorlinks   = true, 
  urlcolor     = blue,
  linkcolor    = blue, 
  citecolor   = blue 
} 

\usepackage{booktabs} 
\usepackage{multirow} 
\usepackage{makecell} 
\usepackage{comment}
\usepackage{amsmath}
\usepackage{xcolor} 
\usepackage{ragged2e}
\usepackage{amsmath} 

\graphicspath{{./}{figures/}}

\begin{document}

\title{exoALMA III: Line-intensity Modeling and System Property Extraction from Protoplanetary Disks}

\author[0000-0001-8446-3026]{Andr\'es F. Izquierdo}
\altaffiliation{NASA Hubble Fellowship Program Sagan Fellow}
\affiliation{Department of Astronomy, University of Florida, Gainesville, FL 32611, USA}
\affiliation{Leiden Observatory, Leiden University, P.O. Box 9513, NL-2300 RA Leiden, The Netherlands}
\affiliation{European Southern Observatory, Karl-Schwarzschild-Str. 2, D-85748 Garching bei M\"unchen, Germany}

\correspondingauthor{Andr\'es F. Izquierdo}
\email{andres.izquierdo.c@gmail.com}


\author[0000-0002-0491-143X]{Jochen Stadler} 
\affiliation{Universit\'e C\^ote d'Azur, Observatoire de la C\^ote d'Azur, CNRS, Laboratoire Lagrange, 06304 Nice, France}

\author[0000-0002-5503-5476]{Maria Galloway-Sprietsma}
\affiliation{Department of Astronomy, University of Florida, Gainesville, FL 32611, USA}

\author[0000-0002-7695-7605]{Myriam Benisty}
\affiliation{Universit\'e C\^ote d'Azur, Observatoire de la C\^ote d'Azur, CNRS, Laboratoire Lagrange, 06304 Nice, France}
\affiliation{Max-Planck Institute for Astronomy (MPIA), Königstuhl 17, 69117 Heidelberg, Germany}

\author[0000-0001-5907-5179]{Christophe Pinte}
\affiliation{Univ. Grenoble Alpes, CNRS, IPAG, 38000 Grenoble, France}
\affiliation{School of Physics and Astronomy, Monash University, Clayton VIC 3800, Australia}

\author[0000-0001-7258-770X]{Jaehan Bae}
\affiliation{Department of Astronomy, University of Florida, Gainesville, FL 32611, USA}

\author[0000-0003-1534-5186]{Richard Teague}
\affiliation{Department of Earth, Atmospheric, and Planetary Sciences, Massachusetts Institute of Technology, Cambridge, MA 02139, USA}

\author[0000-0003-4689-2684]{Stefano Facchini}
\affiliation{Dipartimento di Fisica, Universit\`a degli Studi di Milano, Via Celoria 16, 20133 Milano, Italy}

\author[0000-0002-7212-2416]{Lisa W\"olfer} 
\affiliation{Department of Earth, Atmospheric, and Planetary Sciences, Massachusetts Institute of Technology, Cambridge, MA 02139, USA}

\author[0000-0003-4663-0318]{Cristiano Longarini} 
\affiliation{Institute of Astronomy, University of Cambridge, Madingley Road, CB3 0HA, Cambridge, UK}
\affiliation{Dipartimento di Fisica, Universit\`a degli Studi di Milano, Via Celoria 16, 20133 Milano, Italy}

\author[0000-0003-2045-2154]{Pietro Curone} 
\affiliation{Dipartimento di Fisica, Universit\`a degli Studi di Milano, Via Celoria 16, 20133 Milano, Italy}
\affiliation{Departamento de Astronom\'ia, Universidad de Chile, Camino El Observatorio 1515, Las Condes, Santiago, Chile}


\author[0000-0003-2253-2270]{Sean M. Andrews}
\affiliation{Center for Astrophysics | Harvard \& Smithsonian, Cambridge, MA 02138, USA}

\author[0000-0001-6378-7873]{Marcelo Barraza-Alfaro}
\affiliation{Department of Earth, Atmospheric, and Planetary Sciences, Massachusetts Institute of Technology, Cambridge, MA 02139, USA}

\author[0000-0002-2700-9676]{Gianni Cataldi} 
\affiliation{National Astronomical Observatory of Japan, 2-21-1 Osawa, Mitaka, Tokyo 181-8588, Japan}

\author[0000-0003-3713-8073]{Nicolás Cuello} 
\affiliation{Univ. Grenoble Alpes, CNRS, IPAG, 38000 Grenoble, France}

\author[0000-0002-1483-8811]{Ian Czekala} 
\affiliation{School of Physics \& Astronomy, University of St. Andrews, North Haugh, St. Andrews KY16 9SS, UK}
\affiliation{Centre for Exoplanet Science, University of St. Andrews, North Haugh, St. Andrews, KY16 9SS, UK}

\author[0000-0003-4679-4072]{Daniele Fasano} 
\affiliation{Universit\'e C\^ote d'Azur, Observatoire de la C\^ote d'Azur, CNRS, Laboratoire Lagrange, 06304 Nice, France}

\author[0000-0002-9298-3029]{Mario Flock} 
\affiliation{Max-Planck Institute for Astronomy (MPIA), Königstuhl 17, 69117 Heidelberg, Germany}

\author[0000-0003-1117-9213]{Misato Fukagawa} 
\affiliation{National Astronomical Observatory of Japan, Osawa 2-21-1, Mitaka, Tokyo 181-8588, Japan}

\author[0000-0002-5910-4598]{Himanshi Garg}
\affiliation{School of Physics and Astronomy, Monash University, Clayton VIC 3800, Australia}

\author[0000-0002-8138-0425]{Cassandra Hall} 
\affiliation{Department of Physics and Astronomy, The University of Georgia, Athens, GA 30602, USA}
\affiliation{Center for Simulational Physics, The University of Georgia, Athens, GA 30602, USA}
\affiliation{Institute for Artificial Intelligence, The University of Georgia, Athens, GA, 30602, USA}

\author[0000-0003-1502-4315]{Iain Hammond} 
\affiliation{School of Physics and Astronomy, Monash University, VIC 3800, Australia}

\author[0000-0001-7641-5235]{Thomas Hilder} 
\affiliation{School of Physics and Astronomy, Monash University, VIC 3800, Australia}

\author[0000-0001-6947-6072]{Jane Huang} 
\affiliation{Department of Astronomy, Columbia University, 538 W. 120th Street, Pupin Hall, New York, NY, USA}

\author[0000-0003-1008-1142]{John~D.~Ilee} 
\affiliation{School of Physics and Astronomy, University of Leeds, Leeds, UK, LS2 9JT}

\author[0000-0001-8061-2207]{Andrea Isella}
\affiliation{Department of Physics and Astronomy, Rice University, 6100 Main St, Houston, TX 77005, USA}
\affiliation{Rice Space Institute, Rice University, 6100 Main St, Houston, TX 77005, USA}

\author[0000-0001-7235-2417]{Kazuhiro Kanagawa} 
\affiliation{College of Science, Ibaraki University, 2-1-1 Bunkyo, Mito, Ibaraki 310-8512, Japan}

\author[0000-0002-8896-9435]{Geoffroy Lesur} 
\affiliation{Univ. Grenoble Alpes, CNRS, IPAG, 38000 Grenoble, France}

\author[0000-0002-2357-7692]{Giuseppe Lodato} 
\affiliation{Dipartimento di Fisica, Universit\`a degli Studi di Milano, Via Celoria 16, 20133 Milano, Italy}

\author[0000-0002-8932-1219]{Ryan A. Loomis}
\affiliation{National Radio Astronomy Observatory, 520 Edgemont Rd., Charlottesville, VA 22903, USA}

\author[0000-0003-4039-8933]{Ryuta Orihara} 
\affiliation{College of Science, Ibaraki University, 2-1-1 Bunkyo, Mito, Ibaraki 310-8512, Japan}

\author[0000-0002-4716-4235]{Daniel J. Price} 
\affiliation{School of Physics and Astronomy, Monash University, Clayton VIC 3800, Australia}

\author[0000-0003-4853-5736]{Giovanni Rosotti} 
\affiliation{Dipartimento di Fisica, Universit\`a degli Studi di Milano, Via Celoria 16, 20133 Milano, Italy}

\author[0000-0003-1859-3070]{Leonardo Testi}
\affiliation{Dipartimento di Fisica e Astronomia, Università di Bologna, I-40129 Bologna, Italy}

\author[0000-0003-1412-893X]{Hsi-Wei Yen} 
\affiliation{Academia Sinica Institute of Astronomy \& Astrophysics, 11F of Astronomy-Mathematics Building, AS/NTU, No.1, Sec. 4, Roosevelt Rd, Taipei 10617, Taiwan}

\author[0000-0002-3468-9577]{Gaylor Wafflard-Fernandez} 
\affiliation{Univ. Grenoble Alpes, CNRS, IPAG, 38000 Grenoble, France}

\author[0000-0003-1526-7587	]{David J. Wilner} 
\affiliation{Center for Astrophysics | Harvard \& Smithsonian, Cambridge, MA 02138, USA}

\author[0000-0002-7501-9801]{Andrew J. Winter}
\affiliation{Universit\'e C\^ote d'Azur, Observatoire de la C\^ote d'Azur, CNRS, Laboratoire Lagrange, 06304 Nice, France}
\affiliation{Max-Planck Institute for Astronomy (MPIA), Königstuhl 17, 69117 Heidelberg, Germany}

\author[0000-0001-8002-8473	]{Tomohiro C. Yoshida} 
\affiliation{National Astronomical Observatory of Japan, 2-21-1 Osawa, Mitaka, Tokyo 181-8588, Japan}
\affiliation{Department of Astronomical Science, The Graduate University for Advanced Studies, SOKENDAI, 2-21-1 Osawa, Mitaka, Tokyo 181-8588, Japan}

\author[0000-0001-9319-1296	]{Brianna Zawadzki} 
\affiliation{Department of Astronomy, Van Vleck Observatory, Wesleyan University, 96 Foss Hill Drive, Middletown, CT 06459, USA}
\affiliation{Department of Astronomy \& Astrophysics, 525 Davey Laboratory, The Pennsylvania State University, University Park, PA 16802, USA}

\begin{abstract}

The ALMA large program \exoalma{} offers a unique window into the three-dimensional physical and dynamical properties of 15 circumstellar discs where planets may be actively forming. Here, we present an analysis methodology to map the gas disc structure and substructure encoded in \twCO{}, \thCO{}, and CS line emission from our targets. To model and characterize the disc structure probed by optically thin species, such as CS and, in some cases, \thCO{}, we introduce a composite line profile kernel that accounts for increased intensities caused by the projected overlap between the disc's front and back side emission. 
Our workflow, built on the \discminer{} modelling framework, incorporates an improved iterative two-component fitting method for inclined sources ($i>40^\circ$), to mitigate the impact of the disc backside on the extraction of velocity maps. Also, we report best-fit parameters for the Keplerian stellar masses, as well as inclinations, position angles, systemic velocities, rotation direction, and emission surfaces of the discs in our sample. %

\end{abstract}


\section{Introduction}
\label{sec:intro}

Since the initial reports of candidate planet-driven perturbations in the discs of HD\,163296 \citep{pinte+2018_kink, teague+2018a} and HD\,97048 \citep{pinte+2019}, significant progress has been made in the theoretical and observational understanding of the velocity signatures induced by planets embedded in discs \citep{dong+2019, bollati+2021, bae+2021, rabago+2021, teague+2021, izquierdo+2021, calcino+2022}, and effort has been directed toward recovering these features in a statistically robust manner \citep{izquierdo+2022, izquierdo+2023, stadler+2023}. As part of the first planet-hunting campaign at (sub)mm wavelengths led by the ALMA large program \exoalma{} \citep{Teague_exoALMA}, we build on recent advancements in modelling and analysis of molecular line properties to
investigate the physical and dynamical structure of discs and identify gas substructures potentially sculpted by planets in formation. 

Our methodology features a guided algorithm for performing two-component fits to line profiles, leading to more precise extraction of velocity maps, and the incorporation of channel-map models for optically thin emission. To illustrate our analysis workflow, we use molecular line data from a subset of the \exoalma{} targets, specifically the discs of \hdone{} and \lkca{} as observed in \twCO{}, \thCO{} and CS emission.

The paper is organized as follows. {Section \ref{sec:data} describes the datasets used in this work. Section \ref{sec:models} outlines the methodology for modelling} line intensity channels and presents the best-fit model parameters for the discs in the \exoalma{} sample. Section \ref{sec:observables} demonstrates the extraction of line profile observables and their connection to the physical and dynamical structure of our targets. Finally, Section \ref{sec:conclusions} summarizes the results of this study. 

\section{Data}
\label{sec:data}

The data used in this paper comprises continuum-subtracted Fiducial Images of \twCO{} and \thCOfull{}, and \CSfull{} line emission, with a circular $0.''15$ synthesized beam, as defined in \citet{Teague_exoALMA}. For the source-specific analysis of \lkca{}, High Surface Brightness Sensitivity Images with a circular $0.''3$ synthesized beam are employed to capture the disc's physical properties over a larger radial extent. 
Across all datasets, the channel spacing used here is 100\,m\,s$^{-1}$ for \twCO{} and \thCO{} images, and 200\,m\,s$^{-1}$ for CS. Details on the calibration and imaging procedures are provided in \citet{Loomis_exoALMA}.

\section{Modelling channel maps}
\label{sec:models}

   \begin{figure*}
   \centering
    \includegraphics[width=0.85\textwidth]{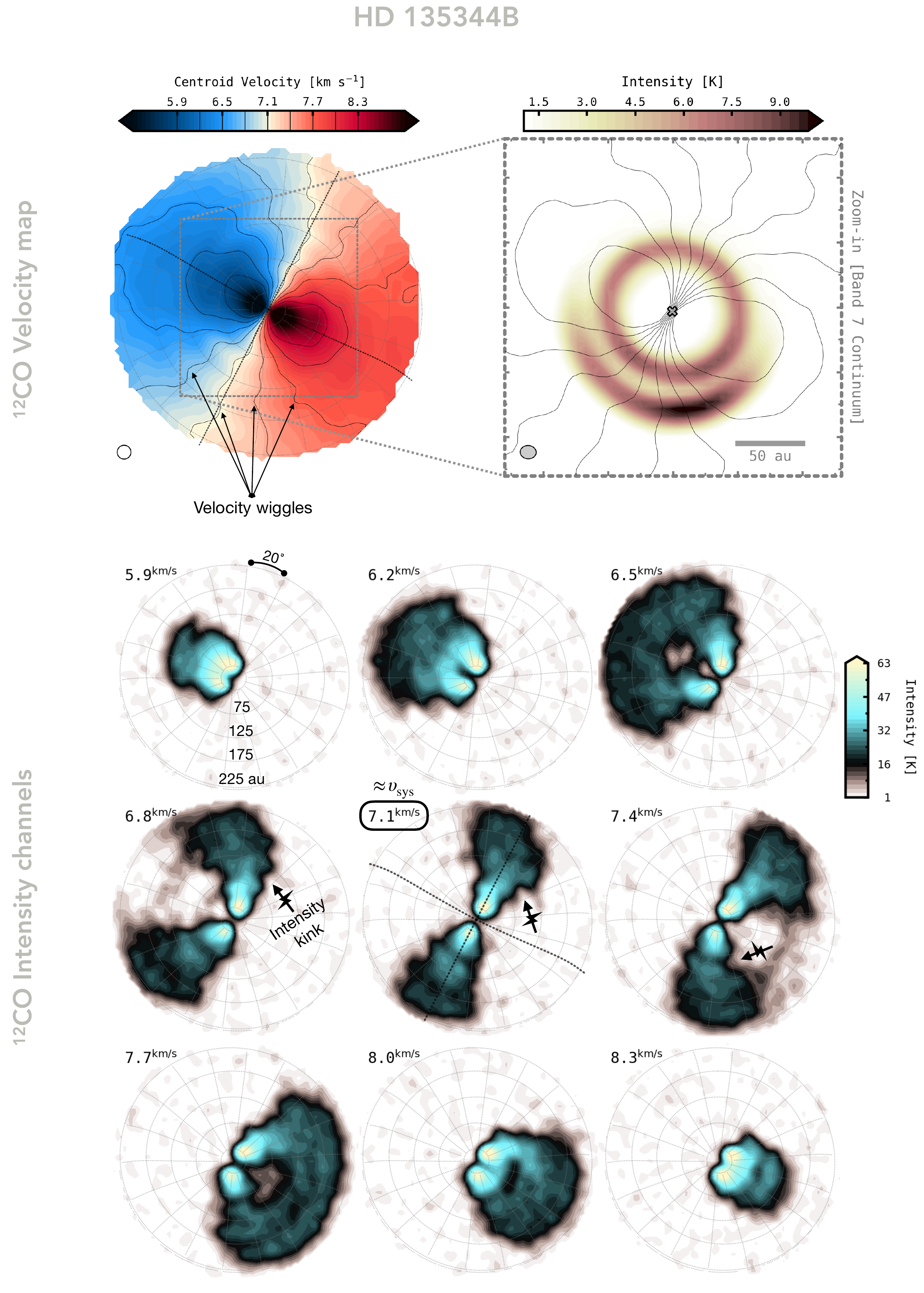}
      \caption{Selected observables from the disc of \hdone{}. \textit{Top row}: Velocity map extracted from the centroid of Gaussian fits to \twCOfull{} line intensity profiles. A zoom-in around the centre of the target illustrates the Band 7 continuum emission and the coexistence of dust substructures and velocity perturbations. \textit{Bottom panels}: Selected \twCO{} intensity channels for the same source, converted to brightness temperature using the Rayleigh-Jeans approximation. In all panels, the background grid follows the \twCO{} emission surface and the best-fit model orientation retrieved by \discminer{}. For reference, arrows mark the location of example intensity kinks identified in the middle row channels \citep[see also][for a discussion]{Pinte_exoALMA}. 
              }
         \label{fig:intro_hd135344_12co}
   \end{figure*}

To study the physical and dynamical structure of the discs targeted by \exoalma{}, we employ the \discminer{} channel-map modelling and analysis framework introduced by \citet{izquierdo+2021}. {In this section, we use our Fiducial datacubes as input for \discminer{} to generate Keplerian model channel maps for \twCO{}, \thCO{}, and CS line emission.}

A high degree of substructure is apparent in most \exoalma{} discs through visual inspection of intensity channels and velocity maps. Figure \ref{fig:intro_hd135344_12co} presents one of our targets, \hdone{}, in \twCO{} emission, where prominent kinks in intensity (bottom panels) and wiggles in velocity (top left) suggest the presence of substantial deviations from Keplerian rotation across much of the disc extent. Interestingly, some of these perturbations coincide with the dusty ring and arc identified in the Band 7 continuum (top right), suggesting a common origin.

To quantify the amplitude of these substructures, we generate smooth \discminer{} models of the disc emission, which can then be subtracted from the data in a subsequent step. Our models adopt a parametric intensity field which, unless otherwise specified, decreases monotonically with radius, and a Keplerian velocity field with differential rotation as a function of height over the midplane. Although this simplified strategy does not account for the underlying physical structure of the disc and radiation transport, it is well-suited for fast parameter exploration and the analysis of large datasets, while maintaining a reasonable degree of realism.

   \begin{figure}
   \centering
    \includegraphics[width=0.8\columnwidth]{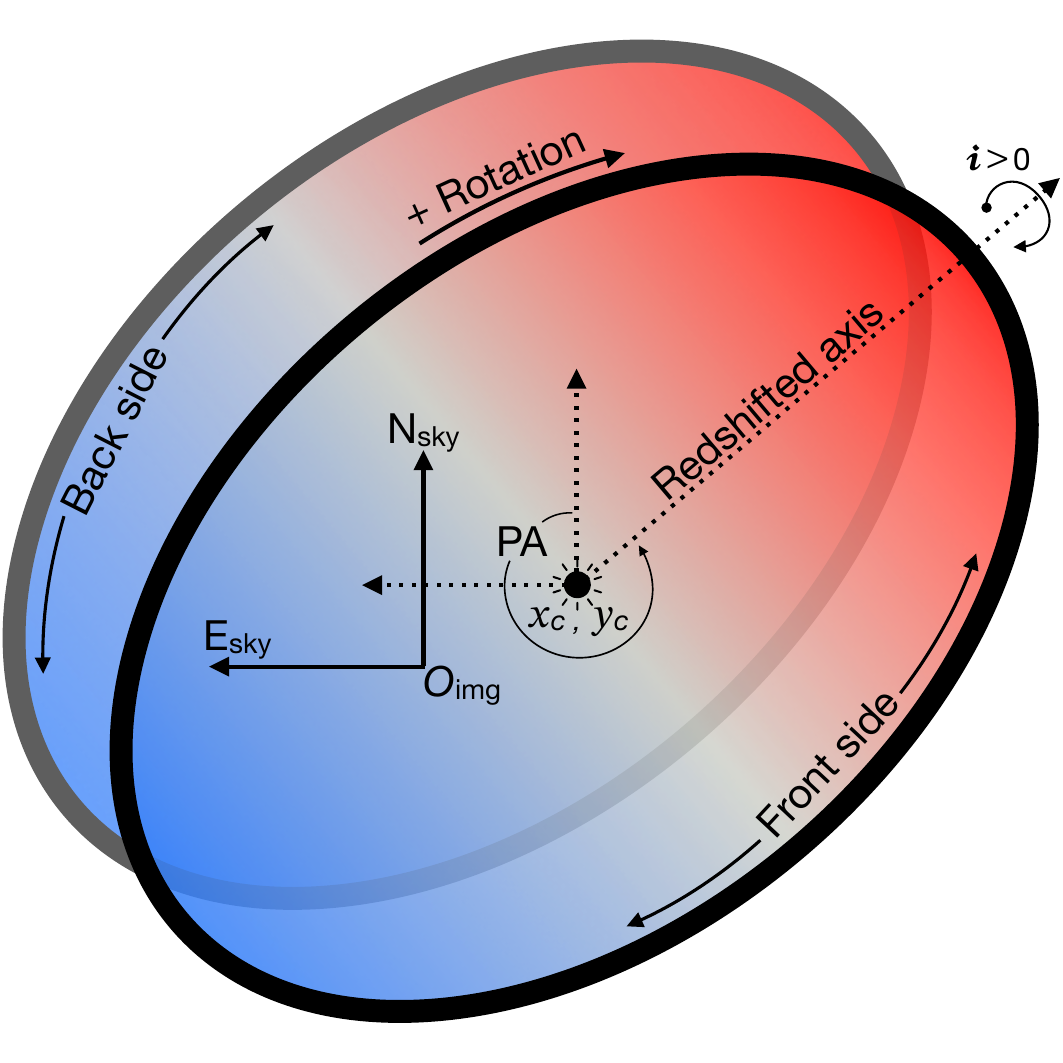}
      \caption{Cartoon illustrating our adopted geometric parameter conventions, and the projected rotation velocity of a disc with positive inclination and clockwise rotation.
              }
         \label{fig:orientation}
   \end{figure}

Our approach incorporates four sets of parameters that control the projected geometry and intensity distribution of the model channels. 
First, orientation parameters that determine the projected appearance of the disc on the sky, including inclination, $i$, position angle, PA, and offsets from the image centre, $x_c$ and $y_c$. {The distance to our sources is fixed to Gaia-derived values as listed in \citet{Teague_exoALMA}.}
Second, velocity parameters that shape the model disc's line-of-sight velocity, that is, Keplerian stellar mass, $M_\star$, systemic velocity, $\upsilon_{\rm LSRK}$, and rotation direction\footnote{
The sign of the inclination and rotation direction are critical factors when modelling sources where both the front and back sides of the disc contribute to the observed intensity. These properties can be determined by locating the red-shifted and blue-shifted sides of the disc relative to the central channel---defined by the systemic velocity---and identifying the half of the disc nearest to the observer (or `near side' for short), which is typically indicated by the prominence of the back side of the disc.}, $\mathrm{sgn}_\mathrm{rot}$. Figure \ref{fig:orientation} illustrates our adopted geometric conventions and the projected appearance of a disc with clockwise rotation {and positive inclination}.

Third, surface parameters that dictate the elevation of the upper and lower emission layers relative to the disc midplane for each molecular line, under the assumption that these are confined to a geometrically narrow region. The observed height of these layers is determined by the line optical depth, which depends on the excitation properties and spatial distribution of the molecular tracer in the gas phase. This distribution is influenced by intrinsic factors, such as the freeze-out temperature of the tracer, and extrinsic factors, including the chemical environment, radiation field, disc temperature, density, and dust content \citep[see][for a review]{miotello+2023}. 

Last, line profile parameters that account for the radial and vertical variations in the shape of the model lines, characterized by the peak intensity, $I_p$, half line width at half power, $L_w$, and line slope, $L_s$. All free parameters and functional forms of the attributes considered by our models are summarized in Table \ref{tab:attributes_parameters}.

\setlength{\tabcolsep}{0.1pt} 

\begin{table}
\centering
{\renewcommand{\arraystretch}{0.9}
\caption{List of default attributes and prescriptions adopted in our \discminer{} models.}  \label{tab:attributes_parameters}

\begin{tabular}{ ll } 

\toprule
\toprule
Attribute & Prescription \\
\midrule


Orientation & $i$, PA, $x_c$, $y_c$ \\ \midrule

Velocity & $\upsilon_{\rm k} = \sqrt{\frac{GM_\star}{r^3}}R$, $\upsilon_{\rm LSRK}$ \\ \midrule

Upper surface & $z = z_{0} (R/D_0)^p \exp{[-(R/R_t)^q]}$ \vspace{0.25cm} \\

Lower surface & $z = -z_{0} (R/D_0)^p \exp{[-(R/R_t)^q]}$ \\

\midrule
a. Peak intensity & $I_p (R \leq R_{\rm out})$ = $I_0 (R/D_0)^p (z/D_0)^q$  \\
    
b. Peak intensity & $I_p (R \leq R_{\rm out})$ = $I_0 (R/R_b)^{p_n} (z/D_0)^q$, \\ $p_n=$ $\begin{cases}
    p_1, & \text{if} \, R \leq R_b \\
    p_2, & \text{if} \, R > R_b 
\end{cases} $

\vspace{0.25cm} \\

Line width & $L_w$ = $L_{w0} (R/D_0)^p (z/D_0)^q$ \vspace{0.25cm} \\ 

Line slope & $L_s$ = $L_{s0} (R/D_0)^p$ \\

\bottomrule

\end{tabular}
 
\justifying
{\noindent \textbf{Note.} $G$ is the gravitational constant; $D_0=100$\,au is a normalisation factor. In disc coordinates, $z$ is the emission height above the disc midplane, $R$ is the cylindrical radius, and $r$ is the spherical radius. The remaining variables are free parameters. All free parameters are modelled independently. The model intensity is set to zero for radii greater than $R_{\rm out}$. Discs with strong intensity gradients near the center are modelled using the type b prescription for the peak intensity. This prescription consists of a broken power-law radial profile characterized by the breaking radius $R_b$.

  }
  
  }
\end{table}

To generate model channel maps, the parametric attributes initially mapped in disc cylindrical coordinates, $(R, \phi)$ and $z(R)$, are first rotated and projected onto the sky plane using the orientation and surface parameters. These on-sky attributes, denoted by primes in the following expression, are then combined into a bell-shaped profile {that independently produces the model intensity, $I_m$, for the upper and lower emission surfaces} as a function of the velocity channel of interest,
\begin{equation} \label{eq:bell}
I_m(\upsilon_{\rm ch}) = I_p'\left(1+\left|\frac{\upsilon_{\rm ch} - \upsilon'_{0}}{L_w'} \right|^{2L_s'} \right)^{-1},\end{equation}
{where $\upsilon'_{0}$ is the model line-of-sight velocity, accounting for the Keplerian motion of the disc at the emission surface $z(R)$, projected onto the sky plane, and referred to the systemic velocity of the source, $\upsilon_{\rm LSRK}$.}

\begin{figure}
   \centering
    \includegraphics[width=1\columnwidth]{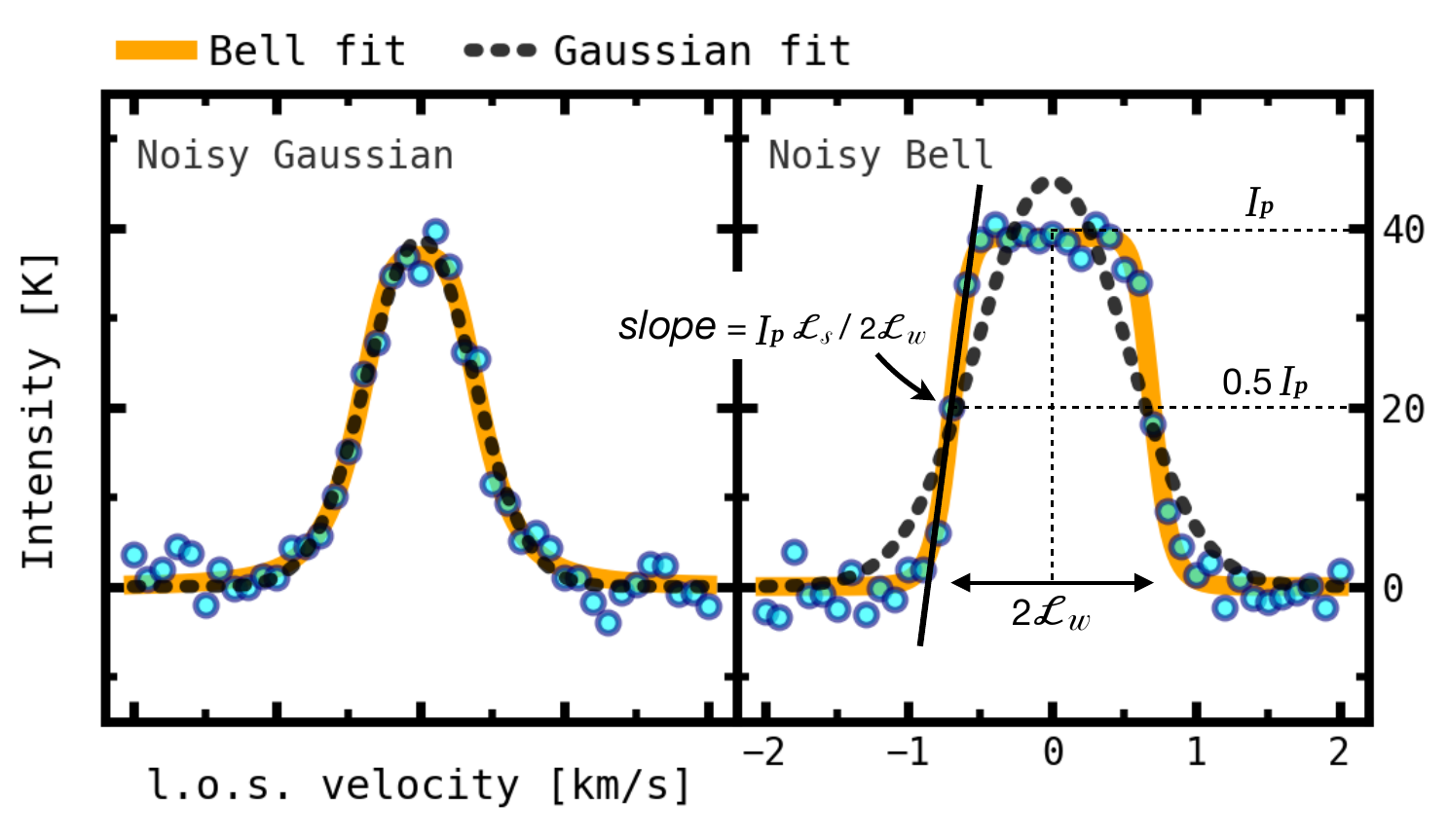}
      \caption{Illustrating Gaussian vs Bell profiles and their corresponding fits. The profiles are sampled by blue circles every 0.1\,\kms{}, with an rms noise of 1.4\,K, {mimicking our Fiducial data \citep[see][]{Teague_exoALMA}.} The true amplitude of both profiles is 40\,K. The Gaussian width at half maximum is 0.41\,\kms{}, which is best fitted by a Bell kernel with an amplitude of $38.4\pm0.9$\,K, a line width of $0.41\pm0.01$\,\kms{}, and a line slope of $1.8\pm0.1$. Thus, while the Bell kernel slightly underestimates the Gaussian amplitude, the retrieved width is consistent.
              }
         \label{fig:bellgauss}
   \end{figure}

The choice of this kernel is motivated by its ability to capture the morphology of optically thick lines, such as those originating from \twCO{} and generally also \thCO{} emission. These lines typically exhibit a flat intensity profile at the peak and a rapid intensity decay towards the wings \citep{hacar+2016}. By introducing one additional parameter---the line slope, $L_s$---the selected kernel outperforms a standard Gaussian in reproducing such profiles. The Bell kernel is also effective for modelling optically thin lines as it approximates a Gaussian profile for low line slopes, as illustrated in Figure \ref{fig:bellgauss}. 

We note that if the optically thin emission is not confined to a geometrically narrow region, the derived surface heights, velocities, and line widths in our model represent an average over multiple disc layers along the line of sight. \citet{paneque-carreno+2024} demonstrated that this layering effect introduces additional non-thermal broadening in optically thin emission from the disc of IM Lup (inclined at $i=47.8^\circ$) by accounting for the radial and vertical Keplerian shear set by a stellar mass of $M_\star=1.1$\,M$_\odot$. The authors found, however, that even with an assumed emission slab as thick as $50$\,au, discrepancies between true and retrieved \discminer{} line widths and centroid velocities remain below 30\,m\,s$^{-1}$. In relation to our data, the optically thinnest of our main tracers, CS, has been suggested to originate from elevations as low as one pressure scale height \citep{dutrey+2017, teague+2022}, which aligns with the typical $z/r$ values independently derived for our sources using \discminer{} and \disksurf{} \citep[see][for a comparison]{Galloway_exoALMA}. Hence, the thickness of the CS emission region is likely smaller than 50\,au across much of the disc extent in most of our targets, minimizing any impact from this layering effect. 


   \begin{figure*}
   \centering
    \includegraphics[width=1\textwidth]{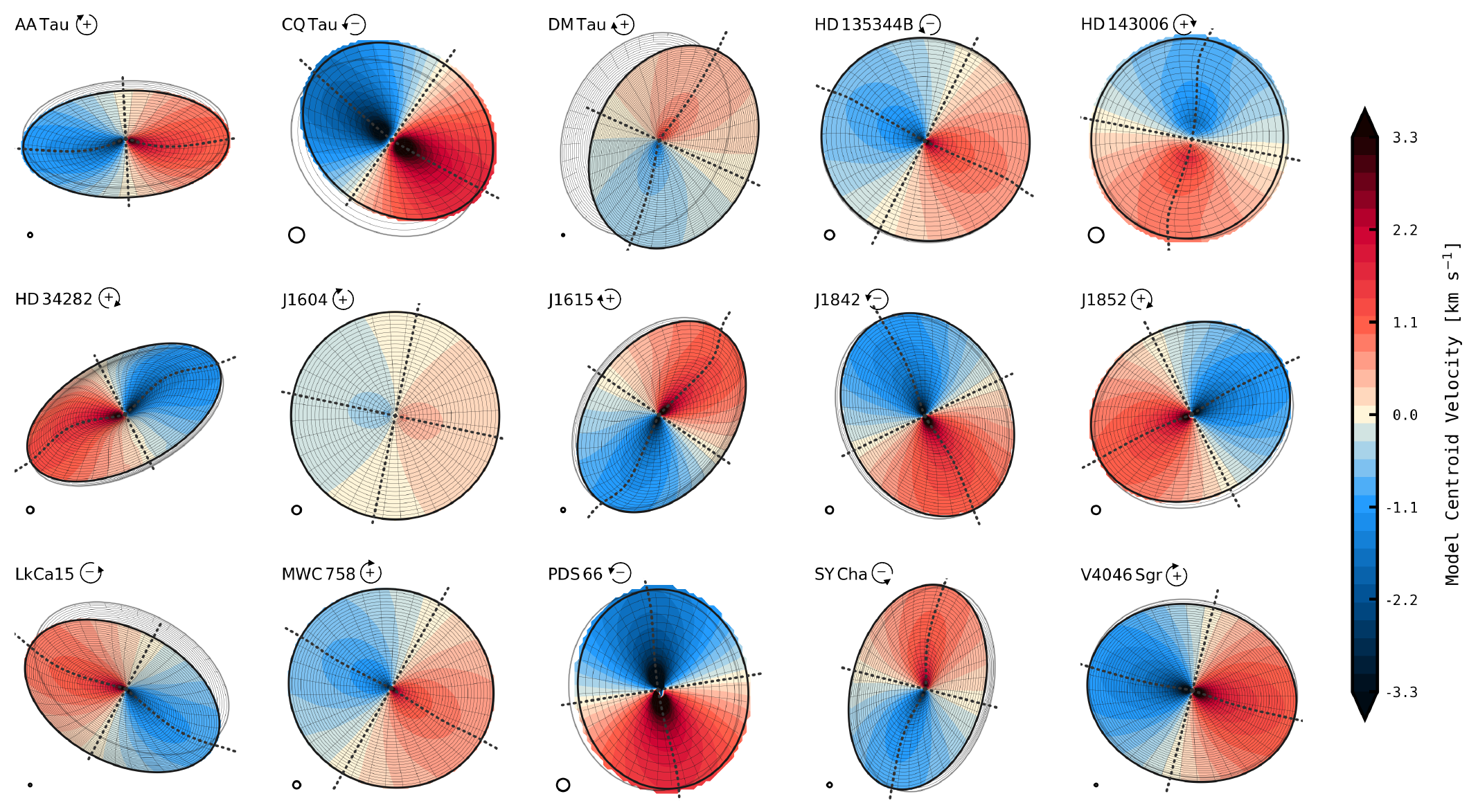}
      \caption{On-sky orientation of the \exoalma{} targets, illustrating the model vertical structure of the discs and the projected Keplerian rotation of their front sides as inferred with \discminer{} {using Fiducial Images of the \twCOfull{} line.} The window size is adjusted to match the physical extent of each disc, while the beam size, shown in the lower-left corner, remains fixed at $0.''15$. The rotation direction adopted for each disc is noted beside the target names. The outer extent of the modelled back-side structure is also visible when it is sufficiently separated from the front-side outer edge.
              }
         \label{fig:gallery_velocity}
   \end{figure*}

   \begin{figure*}
   \centering
    \includegraphics[width=1\textwidth]{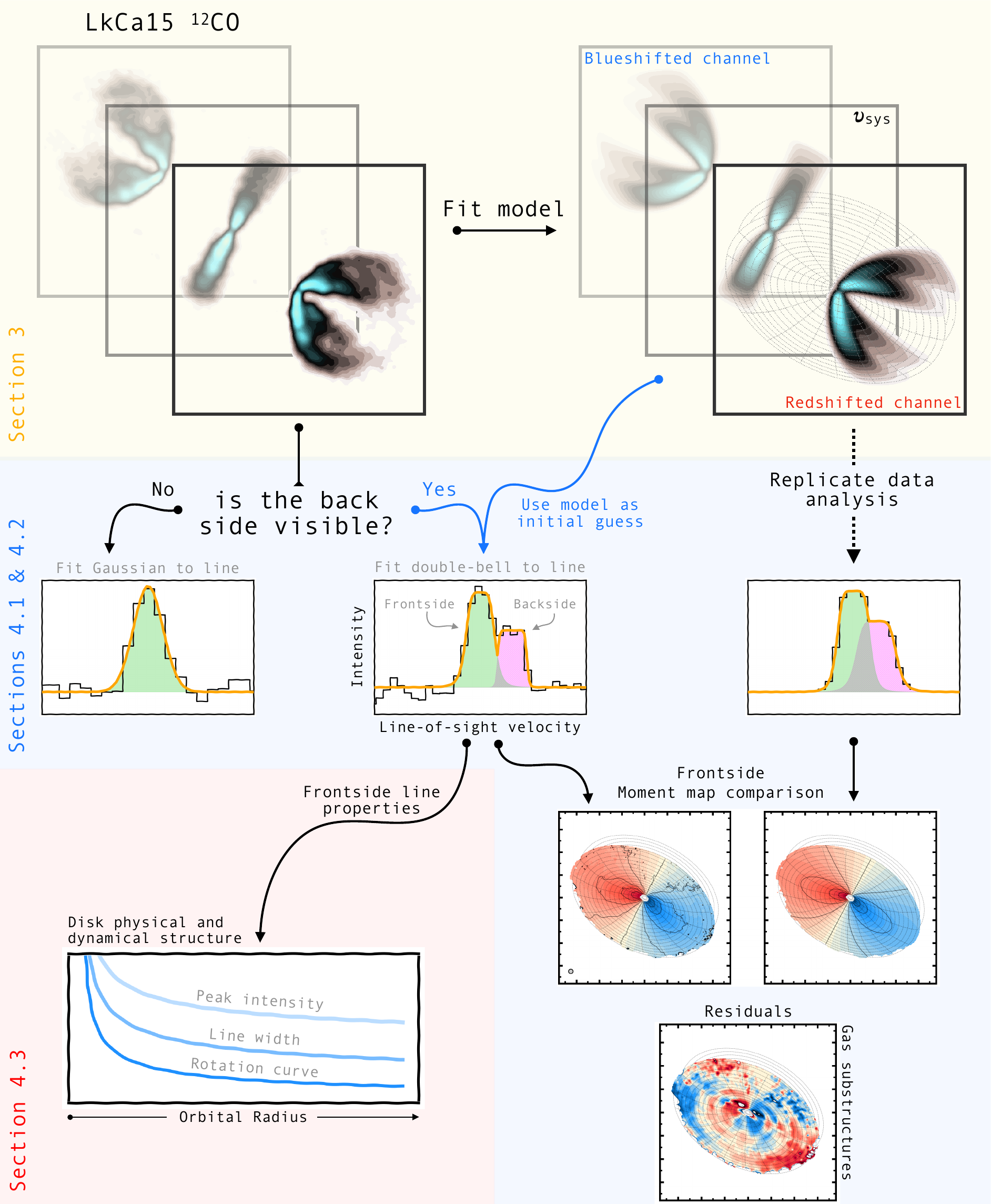}
      \caption{Flowchart illustrating the extraction of line profile properties for \twCOfull{} emission from the disc of \lkca{}. 
              }
         \label{fig:flowchart}
   \end{figure*}

To combine the contributions from the front and back sides of the disc to the total line profile, we adopt two strategies depending on the assumed optical depth of the tracer being analyzed. For optically thick emission, we continue using the method introduced in \citet{izquierdo+2021}, which selects the higher intensity between the two components at each pixel and velocity channel. This results in front-side emission dominating when the front- and back-side line profiles are not sufficiently separated in velocity because of projection. For optically thin tracers, however, we now adopt a different approach. In this case, we sum the front and back side contributions directly to form the total line profile. By employing these two methods, our models can capture representative features of both optically thick and thin emission, including radial and azimuthal intensity variations arising from overlapping column densities along different lines of sight. Since investigating small-scale changes in the discs' optical depth is beyond the scope of this paper, we assume either fully optically thick or fully optically thin emission across the entire disc, rather than a combination of both. A mixed assumption would require a transition through a marginally optically thick state, introducing additional free parameters. The impact of using one approach over the other will be illustrated in Sect. \ref{sec:observables}. 

The modelling library of \discminer{} integrates the \textsc{emcee} ensemble sampler for Markov Chain Monte Carlo (MCMC) \citep{foreman+2013} to identify the optimal combination of parameters that best matches the observed line intensity of the target disc, $I_d$, by maximizing the log-likelihood function,
\begin{equation} \label{eq:loglikelihood}
 \log \mathcal{L} = -0.5\sum_{j}^{n_{\rm ch}} \sum_{i}^{n_{\rm pix}}  \left[I_m(r_i, \upsilon_j) - I_d(r_i, \upsilon_j)\right]^2/\sigma_{i}^{2} ,    
\end{equation}
where $i$ indexes spatial pixels, and $j$ indexes velocity channels. The weighting factor, $\sigma_i$, is taken as the standard deviation of the observed intensity per pixel, measured from line-free velocity channels.

To sample the parameter space, we adopt flat priors and employ a number of walkers at least 10 times the number of free parameters. These walkers are evolved over an average of 20,000 iterations, allowing them to stabilize around a set of optimal parameters, defined as the median values of the last 10\,\% of the iterations. 
Tables \ref{tab:mainpars} and \ref{tab:surfacepars} summarize the best-fit Keplerian stellar masses, systemic velocities, orientation, and surface parameters derived for all \exoalma{} discs from the Fiducial Images introduced in \citet{Teague_exoALMA}. {We refer the reader to \citet{Longarini_exoALMA} for a summary of the dynamical stellar masses derived from our sources, considering disc pressure and self-gravity forces.} In Figure \ref{fig:gallery_velocity}, we provide a gallery of the projected vertical structure and rotation velocity of the targets as inferred by our modelling procedure.
Parameter uncertainties derived from our posterior distributions for a mid-inclination source like \lkca{} ($i=50.4^\circ$), are around $0.1\%$ for the stellar mass and inclination, and $0.5\%$ for the upper-surface height normalization, $z_0$. However, our simplified treatment of noise, which neglects correlations between nearby pixels, generally results in underestimated uncertainties. The impact of considering spatial correlations in the noise on our posterior distributions is quantitatively explored in \citet{Hilder_exoALMA}, where more realistic uncertainties are found to increase by a factor of $\sim\!10$ for velocity, orientation, and surface parameters.
For illustration, Figures \ref{fig:channel_maps_hd135344} and \ref{fig:channel_maps_lkca15} in Appendix \ref{sec:supporting_figures}, present selected intensity channels of \twCO{}, \thCO{}, and CS from our observations, along with the best-fit models and the corresponding residuals.

\setlength{\tabcolsep}{6.5pt} 

\begin{table*}
\centering
{\renewcommand{\arraystretch}{0.95}
\caption{Best-fit Keplerian stellar mass, systemic velocity, orientation parameters, and rotation direction inferred for our sources from Fiducial Images of \twCOfull{} line emission, which have a beam size of $0.''15$ and a channel spacing of 100\,m\,s$^{-1}$ \citep[see][for details on the calibration and imaging procedures]{Teague_exoALMA, Loomis_exoALMA}. Also listed are the fiducial moment map types used for extracting physical properties in this and other papers in the \exoalma{} series (see Sect. \ref{sec:observables}). Parameter uncertainties are on the order of $0.1\%$. However, \citet{Hilder_exoALMA} found that spatially correlated noise can increase these by a factor of ${\sim}\!10$. Moment map abbreviations: DB=double-bell, G=Gaussian, B=Bell.} \label{tab:mainpars}
\begin{tabular}{|l|c|r|c|rrrr|ccc|}
\hline
\multirow{2}{*}{Target} & \multicolumn{1}{c|}{$M_{\star}$} & \multicolumn{1}{c|}{$\upsilon_{\rm LSRK}$} & {Rotation} & \multicolumn{1}{c}{$i$} & \multicolumn{1}{c}{PA} & \multicolumn{1}{c}{$x_c$} & \multicolumn{1}{c|}{$y_c$} & \multicolumn{3}{c|}{Moment map type} \\
\cline{9-11}
         & $({\rm M}_{\odot})$ & (km\,s$^{-1}$) & \( \stackrel{\displaystyle \circlearrowright}{\scriptstyle +} \)  \( \stackrel{\displaystyle \circlearrowleft}{\scriptstyle -} \) & \multicolumn{1}{c}{(deg)} & \multicolumn{1}{c}{(deg)} & \multicolumn{1}{c}{(mas)} & \multicolumn{1}{c|}{(mas)} & \twCO{} & \thCO{} & CS \\
\hline
\dmtau{}    & 0.45 & 6.03 & $+$ & $39.8$  & 335.7 & $-71.2$ & $-31.4$ & DB & G  & G \\
\aatau{}    & 0.79 & 6.50 & $+$ & $-58.7$ & 272.7 & $-25.6$ & 47.6 & DB & DB & G  \\
\lkca{}   & 1.17 & 6.29 & $-$ & $50.4$ & $61.9$  & $-41.5$ & 50.2 & DB & DB & G \\
\hdthree{}  & 1.62 & $-2.33$ & $+$ & $-58.3$ & 117.4 & $-12.7$ & 25.3 & DB & DB & G \\
\mwcsev{}   & 1.40  & 5.89 & $+$ & $19.4$  & $240.3$  & $-5.3$ & 48.7 & G  & G  & G \\
\cqtau{}    & 1.40  & 6.19 & $-$ & $-36.3$ & 235.1 & $-21.5$ & 19.4 & G  & G  & G \\
\sycha{}    & 0.81 & 4.10 & $-$ & $-50.7$ & $345.6$ & $-27.7$ & 37.3 & DB & DB & G \\
\pdssix{}    & 1.28 & 3.96 & $-$ & $-31.9$ & $189.0$ & $-20.2$ & 19.2 & G  & G  & G \\
\hdone{} & 1.61 & 7.09 & $-$ & $-16.1$ & $242.9$ & $-21.0$ & 1.8 & G  & G  & G \\
\hdonefour{} & 1.56 & 7.72 & $+$ & $-16.9$ & 168.2 & $-29.4$ & 26.5 & G  & G  & G \\
\jfour{}    & 1.29 & 4.62 & $+$ & $6.0$   & 258.1 & $-78.3$ & 2.3 & B  & B  & G \\
\jfifteen{}    & 1.14 & 4.75 & $+$ & $46.1$  & 325.4 & $-35.9$ & 9.6 & DB & DB & G \\
\vforty{}    & 1.76 & 2.93 & $+$ & $-33.6$ & $255.7$ & $-64.1$ & $-38.6$ & G  & G  & G \\
\jfortytwo{}    & 1.07 & 5.94 & $-$ & $39.4$  & 205.9 & $-19.6$ & $-8.0$ & DB & G  & G \\
\jfiftytwo{}    & 1.03 & 5.47 & $+$ & $-32.7$ & 117.1 & $-30.1$ & 27.1 & G  & G  & G \\
\hline
\end{tabular}
}
\end{table*}

\setlength{\tabcolsep}{3.5pt} 

\begin{table*}
    \centering
    {\renewcommand{\arraystretch}{0.95}
    \caption{Best-fit model emission surface parameters {inferred for our sources from Fiducial Images of \twCO{}, \thCO{}, and CS. The orientation parameters and systemic velocity of the \thCO{} and CS models were fixed to the \twCO{} values reported in Table \ref{tab:mainpars}.} The ``Front+Back side'' column indicates how the front and back side contributions were combined into a composite model line profile, as detailed in Sect. \ref{sec:models}. For optically thick emission, we use `mask'; for optically thin, we use `sum'. Blank spaces mark models where the back side does not affect the disc emission symmetry, either due to a near face-on orientation or a shallow vertical structure (see Figures \ref{fig:flowchart} through \ref{fig:centroids_lkca15_12co}).}  \label{tab:surfacepars}
    \begin{tabular}{|l|ccc|ccc|ccc|ccc|ccc|}
        \hline
        \multirow{2}{*}{Target} & \multicolumn{3}{c|}{Front + Back side} & \multicolumn{3}{c|}{$z_0$\,(au)} & \multicolumn{3}{c|}{$p$} & \multicolumn{3}{c|}{$R_t$\,(au)} & \multicolumn{3}{c|}{$q$} \\
        \cline{2-16}
        & \twCO{} & \thCO{} & CS & \twCO{} & \thCO{} & CS & \twCO{} & \thCO{} & CS & \twCO{} & \thCO{} & CS & \twCO{} & \thCO{} & CS \\
        \hline
        \dmtau{} & mask & mask & sum & 86.6 & 21.5 & 6.5  & 1.87 & 2.01  & 3.17  & 80   & 295 & 94  & 0.48  & 1.10   & 0.92  \\
        \aatau{} & mask & mask & sum & 49.8  & 51.7 & 32.9 & 1.20  & 1.36  & 0.83  & 240 & 151 & 232 & 1.35  & 1.35  & 2.18  \\
        \lkca{} & mask & sum  & sum & 29.0 & 27.3 & 29.3 & 1.06 & 0.87  & 0.72  & 795 & 511 & 303 & 3.19  & 3.46  & 4.56  \\
        \hdthree{} & mask & mask & sum & 34.0 & 27.2 & 15.9 & 1.19 & 0.79  & 1.72  & 512 & 510 & 268 & 3.20   & 4.41  & 1.82  \\
        \mwcsev{} & mask & -- & -- & 16.3 & 66.5 & 7.6  & 0.97 & 3.22  & 4.96  & 254 & 12  & 111 & 5.34  & 0.71  & 3.29  \\
        \cqtau{} & mask & -- & -- & 41.7 & 38.8 & 28.0 & 1.25 & 1.09  & 4.59  & 346 & 17  & 80  & 0.09  & 0.41  & 1.53  \\
        \sycha{} & mask & sum  & sum & 43.4 & 72.9 & 49.7 & 1.79 & 2.44  & 1.92  & 210  & 66 & 124 & 1.02  & 0.70   & 1.75  \\
        \pdssix{} & mask & -- & -- & 17.4 & 7.5  & 1.2  & 1.83 & 1.20   & 2.93  & 127 & 29  & 92 & 4.48  & 1.54  & 8.39  \\
        \hdone{} & -- & -- & -- & 13.7 & 10.0  & 0.0   & 1.42 & 1.27  & --   & 226 & 175 & --    & 10.0  & 9.67  & --   \\
        \hdonefour{} & -- & -- & -- & 40.5 & 23.8 & 16.9  & 1.89 & 2.20   & 0.72  & 161 & 146 & 103 & 5.98  & 13.07 & 4.51  \\
        \jfour{} & -- & -- & -- & 0.0   & 0.0   & 0.0   & --  & --   & --   & --    & --    & --    & --   & --   & --   \\
        \jfifteen{} & mask & mask & sum & 26.3 & 19.0 & 37.9  & 1.04 & 1.04  & 2.22  & 530  & 425  & 121 & 6.89  & 5.92  & 0.80   \\
        \vforty{} & mask & mask & -- & 25.9 & 33.5 & 0.1  & 1.84 & 1.57  & 0.0   & 151 & 66 & 86 & 1.17  & 1.14  & 7.07 \\
        \jfortytwo{} & mask & sum  & sum & 25.9 & 17.5 & 28.6 & 1.46 & 1.70   & 1.43  & 211 & 143 & 145 & 1.89  & 2.01  & 4.46  \\
        \jfiftytwo{} & mask & mask & sum & 75.3 & 31.2 & 1.0  & 1.78 & 2.74  & 3.64  & 61 & 90 & 108 & 0.84  & 1.33  & 1.42  \\
        \hline
    \end{tabular}
    }
\end{table*}

\section{Extraction of moment maps and radial profiles from line properties}
\label{sec:observables}

To identify and quantify perturbations in the physical and dynamical structure of our target discs, we focus on the analysis of variations in the morphological properties of the observed molecular lines, compared to those predicted by the smooth and Keplerian models produced with \discminer{}. {In this section, we use the best-fit geometry and velocity parameters derived in Section \ref{sec:models} to demonstrate the extraction of moment maps and radial profiles. The characterization and interpretation of moment map residuals will be presented in upcoming papers in the \exoalma{} series.}

The morphology of molecular lines emitted by discs, and captured by ALMA, can be complex, particularly when the target is vertically extended and inclined with respect to the observer \citep[see][for a review]{pinte+2023}. This spectral complexity arises primarily from the combined contribution of front- and back-side emission along multiple sight lines and can be further influenced by physical and chemical factors, such as attenuation due to dust content \citep{isella+2018} and velocity-selective absorption through the disc midplane \citep{pinte+2018_surfaces, dullemond+2020}. However, this poses a challenge for conventional methods that rely on single-component profiles to extract molecular line properties and produce moment maps.

Hence, we explore an alternative approach to better characterize the lines targeted by \exoalma{}. In this section, we demonstrate the necessity of employing two types of extraction methods: single-component and double-component fits applied to the observed and model line profiles. 
The choice between methods is guided by visual inspection of intensity channels and spectra, and is strongly influenced by the disc inclination, emission surface elevation, and the optical depth of the tracer, as these factors dictate how much of the back-side emission contributes to the total line profile. Figure \ref{fig:flowchart} presents a decision tree illustrating this process for the disc of \lkca{} in \twCO{}, and Table \ref{tab:surfacepars} summarizes our choices for all discs and tracers. {For our line profile fits in this step, we use the Levenberg-Marquardt algorithm for least squares curve fitting, as implemented in the \textsc{curve\_fit} module of the \textsc{scipy.optimize} library.}

   \begin{figure*}
   \centering
    \includegraphics[width=0.75\textwidth]{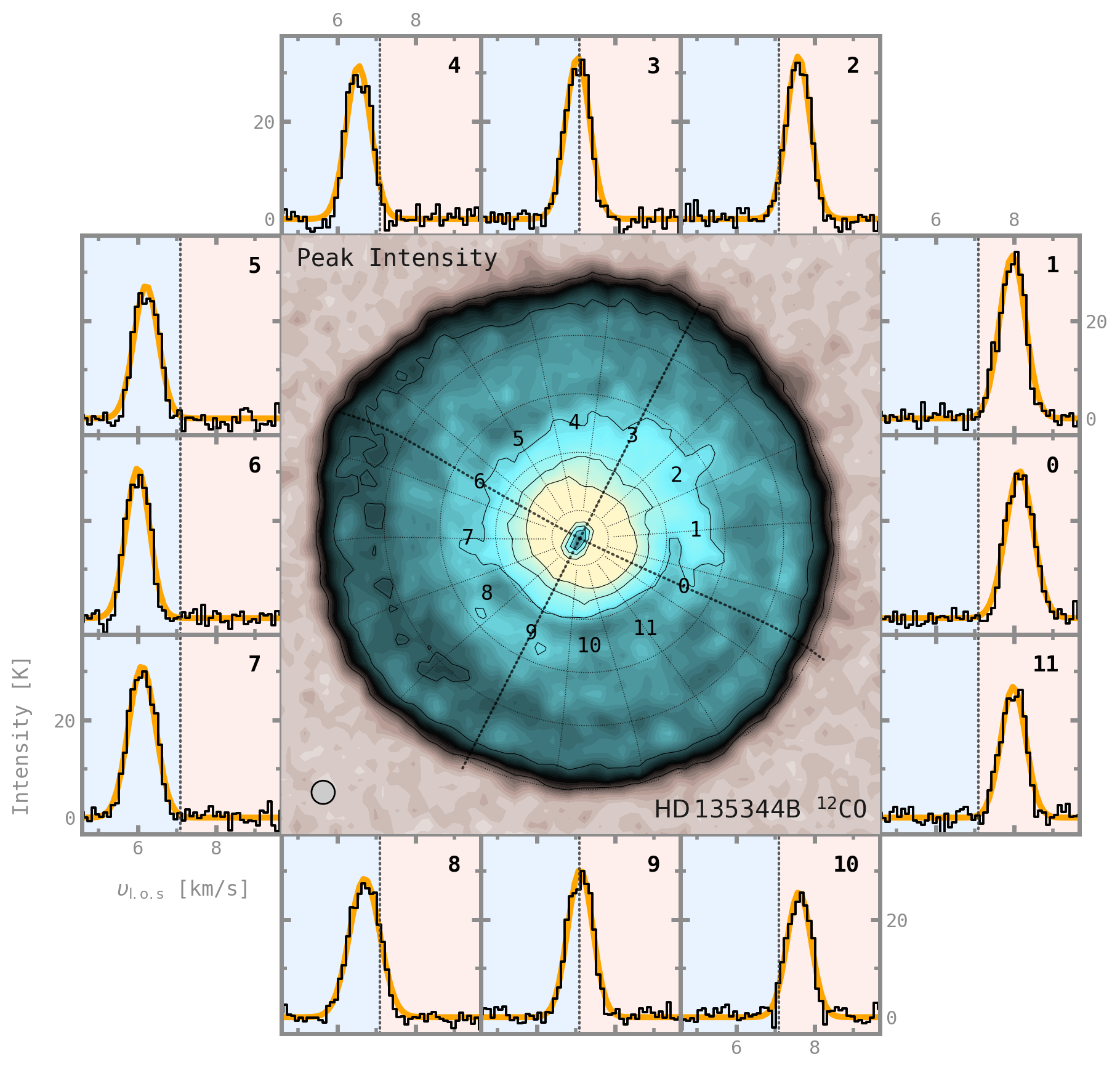}
      \caption{Selected \twCOfull{} spectra extracted from the disc of \hdone{} along an annulus at $R=100$\,au, sampled every 30$^\circ$ in the disc frame, at the locations marked by numbers in the central panel, which displays the peak intensity of the disc. The dotted lines in the outer subpanels highlight the systemic velocity of the object. The black solid lines represent the observed spectra, while the orange lines depict the Gaussian fits applied to them. The background of the subpanels is color-coded, with red and blue regions indicating the redshifted and blueshifted zones, respectively. 
              }
         \label{fig:line_profiles_hd135344_12co}
   \end{figure*}

\subsection{Single-peaked sources} \label{sec:single_moments}

The vertical structure of flat or low inclination discs (i.e. closer to face-on than to edge-on with respect to the observer) is inherently spatially unresolved. Thus, the line profiles emitted by such sources typically exhibit a single peak, making them suitable for fitting with a single-component function like a Gaussian or a Bell profile, 
as demonstrated in Figure \ref{fig:line_profiles_hd135344_12co} for the \twCO{} disc of \hdone{}. This constitutes our default approach for deriving moment maps from these specific targets. Among the 15 discs in our sample, and subject to the angular and velocity resolution of our observations, we note that eight of them fall within this category, all with inclinations smaller than 40$^\circ$.

The channel maps of a flat single-peaked source could be reproduced by a \discminer{} model using a single emitting surface, the front side of the disc. However, if the target line is optically thin, and the source is sufficiently inclined, this prescription does not necessarily lead to a more accurate representation of the disc intensity and velocity field compared to a two-surface model.  
In this scenario, even though the observed line profiles may appear single-peaked, the contribution from the back side of the disc remains active, and its strength varies from one line of sight to another. This additional emission component becomes evident only after analysis of moment maps, and leads to increased peak intensities along the disc minor and major axes, where emission from the front and back sides of the disc adds up due to their proximity in velocity space, and significant enhancements in line widths around the disc diagonal axes, where the overlapping front and back sides have the greatest separation in velocity due to projection. 

To address this complexity, as explained in Sect. \ref{sec:models}, the \discminer{} models of CS emission from mid-- and high-inclination sources incorporate two surfaces that sum in intensity to emulate the effect of overlapping column densities from the front and back sides of the disc. This approach yields a more accurate reproduction of azimuthal features observed in peak intensity and line width maps. Figure \ref{fig:linewidth_peakint_lkca15_cs} in Appendix \ref{sec:supporting_figures} demonstrates this for CS emission from the disc of \lkca{}, where a distinct cross pattern appears in the line width map, and elongated intensity patterns become more pronounced along the disc's major and minor axes. The excellent agreement between the `Front+Back' side model and the observed signatures suggests that these features are unlikely to be solely driven by actual density and temperature fluctuations. Instead, they can be attributed to the effects of finite angular resolution and varying optical depth along different sight lines, {resulting from the overlapping contributions of front- and back-side emission. This effect is also evident in individual intensity channels, as illustrated in Fig. \ref{fig:channel_maps_lkca15}, where regions of increased intensity appear where front- and back-side isovelocities intersect or are closely spaced.}

\subsection{Double-peaked sources} \label{sec:double_moments}

\begin{figure*}
   \centering
    \includegraphics[width=0.497\textwidth]{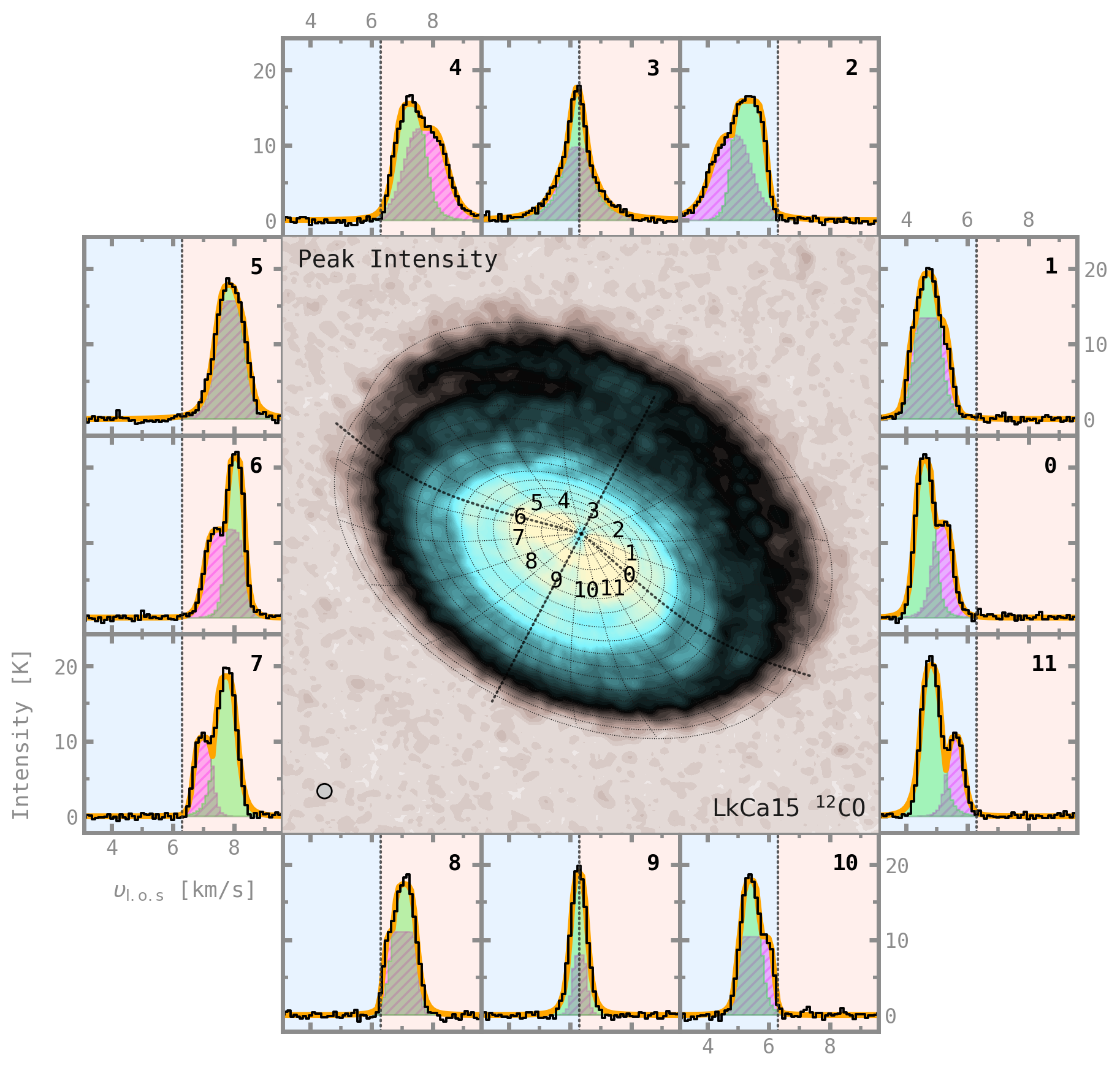}
    \includegraphics[width=0.497\textwidth]{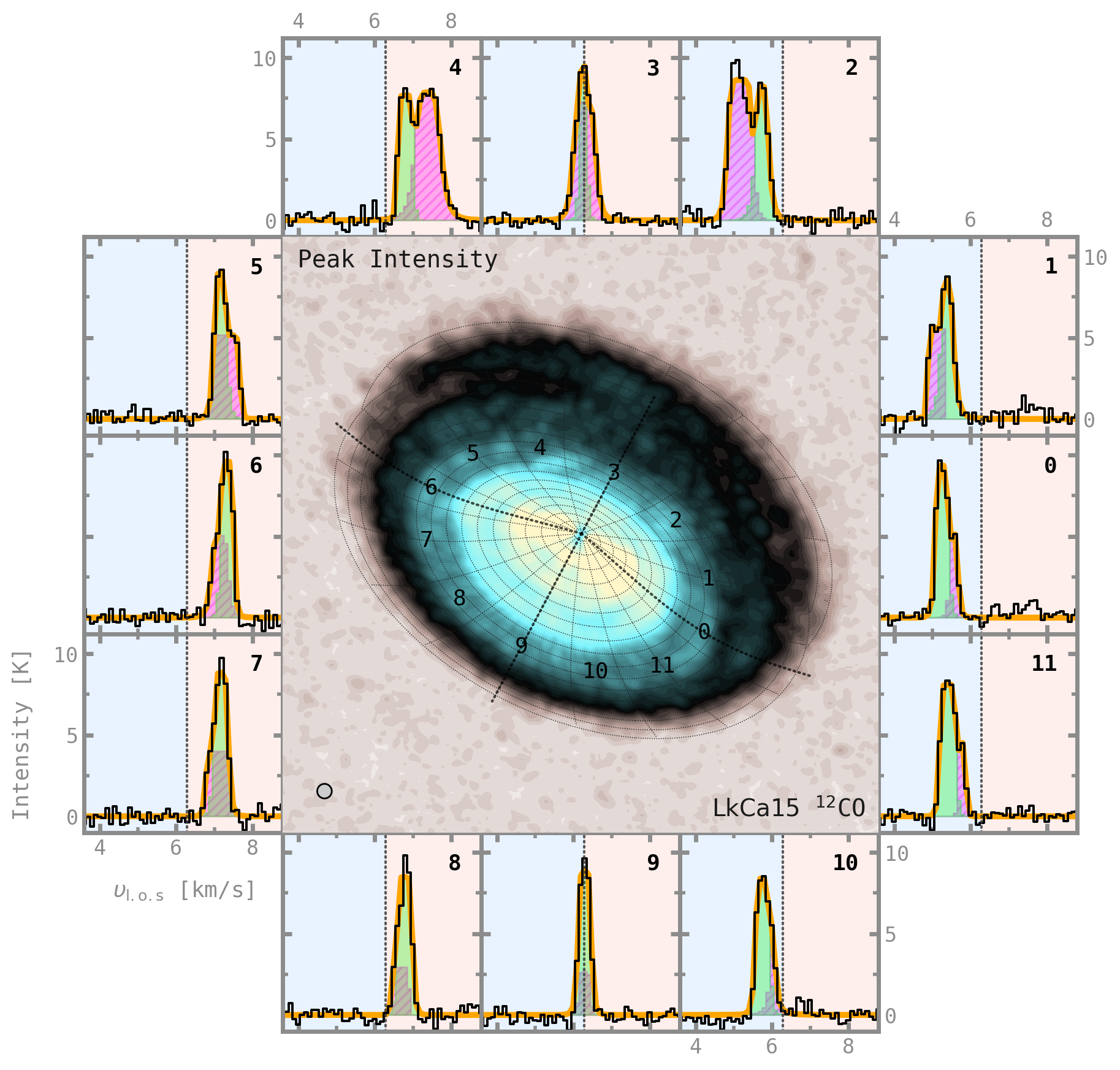}
      \caption{Selected \twCOfull{} spectra from the disc of \lkca{} along annuli at $R=200$\,au (left) and $R=500$\,au (right), sampled every 30$^\circ$ in the disc frame, at the locations marked by numbers in the central panel. The dotted lines in the outer subpanels highlight the systemic velocity of the object. The black solid lines represent the observed spectra, while the orange lines depict the double-bell fits applied to them. Green and magenta shades are also shown within the double-bell fits to illustrate the primary and secondary components of the intensity profile, associated with the front and back sides of the disc, respectively. The background of the subpanels is color-coded, with red and blue regions indicating the redshifted and blueshifted zones.
                    }
         \label{fig:line_profiles_lkca15_12co}
   \end{figure*}

Mid-- and high-inclination discs ($i>40^\circ$) with line emission originating from elevated surfaces generally exhibit double-peaked spectra. To extract moment maps from these sources, we adopt a default approach that involves fitting a double-bell kernel 
to the observed line profiles. This fitting procedure is demonstrated in Figure \ref{fig:line_profiles_lkca15_12co} for the \twCO{} disc around \lkca{}, where line profiles are decomposed into primary (green) and secondary (hatched magenta) components, representing the upper and lower surface contributions to the total intensity, respectively. Unless otherwise specified, we default to using the upper surface component of this fit to report any subsequent quantities derived from these sources.

   \begin{figure*}
   \centering
    \includegraphics[width=0.95\textwidth]{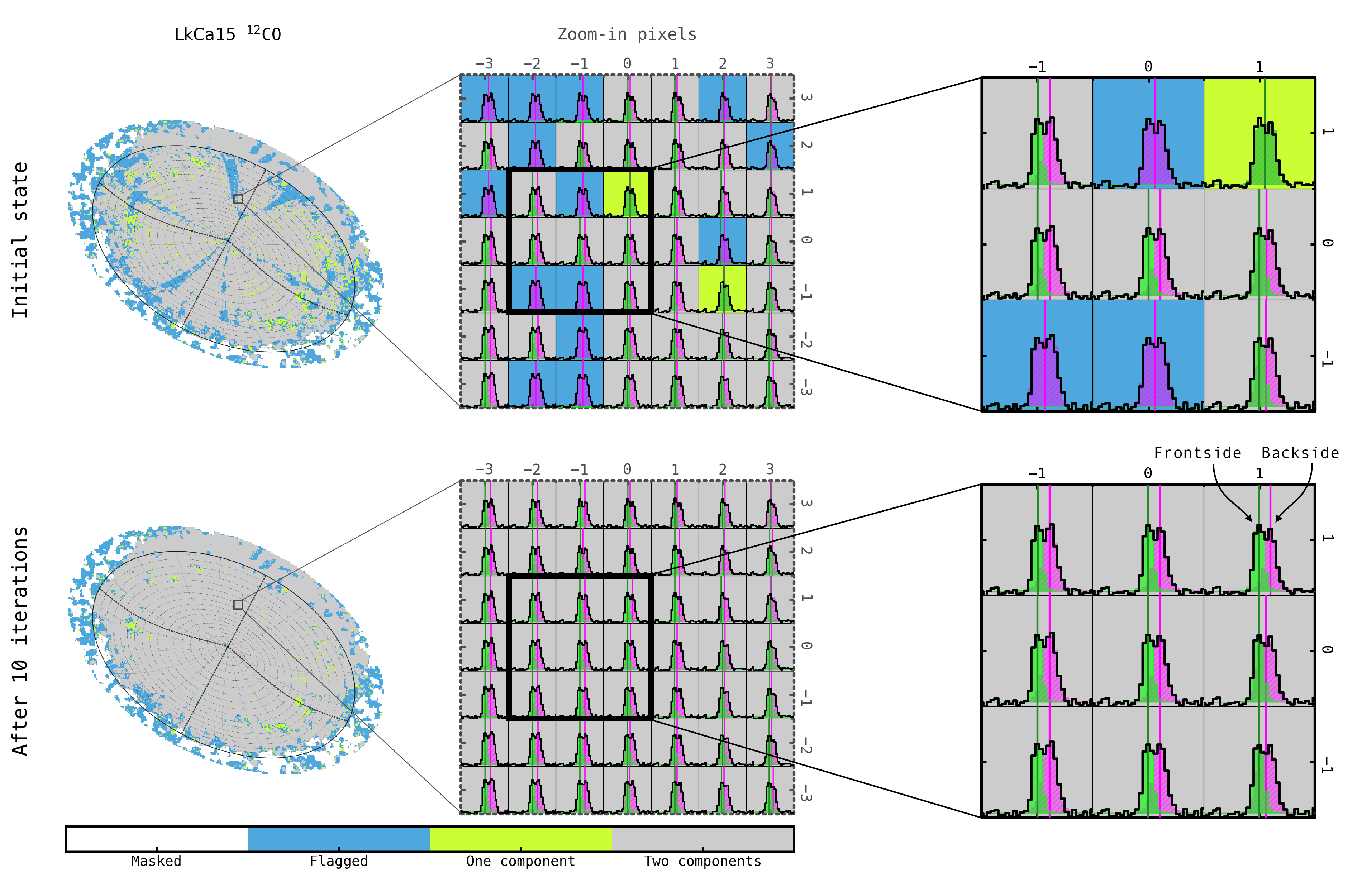}
      \caption{{Illustrating the two-component fitting process for a region of the \lkca{} disc where the back side \twCO{} emission is prominent and sometimes dominant. Pixels colored in blue and yellow are re-fitted using new initial guesses, calculated as the median values of parameters from an $11\times11$ grid of neighboring pixels where the fit was otherwise successful (see Sect. \ref{sec:double_moments}). After 10 iterations, the majority of the originally flagged and single-component pixels have converged into a double-component fit, with well-distinguished front and back side contributions. }
              }
         \label{fig:pixels_lkca15_12co}
   \end{figure*}

   \begin{figure*}
   \centering
    \includegraphics[width=1\textwidth]{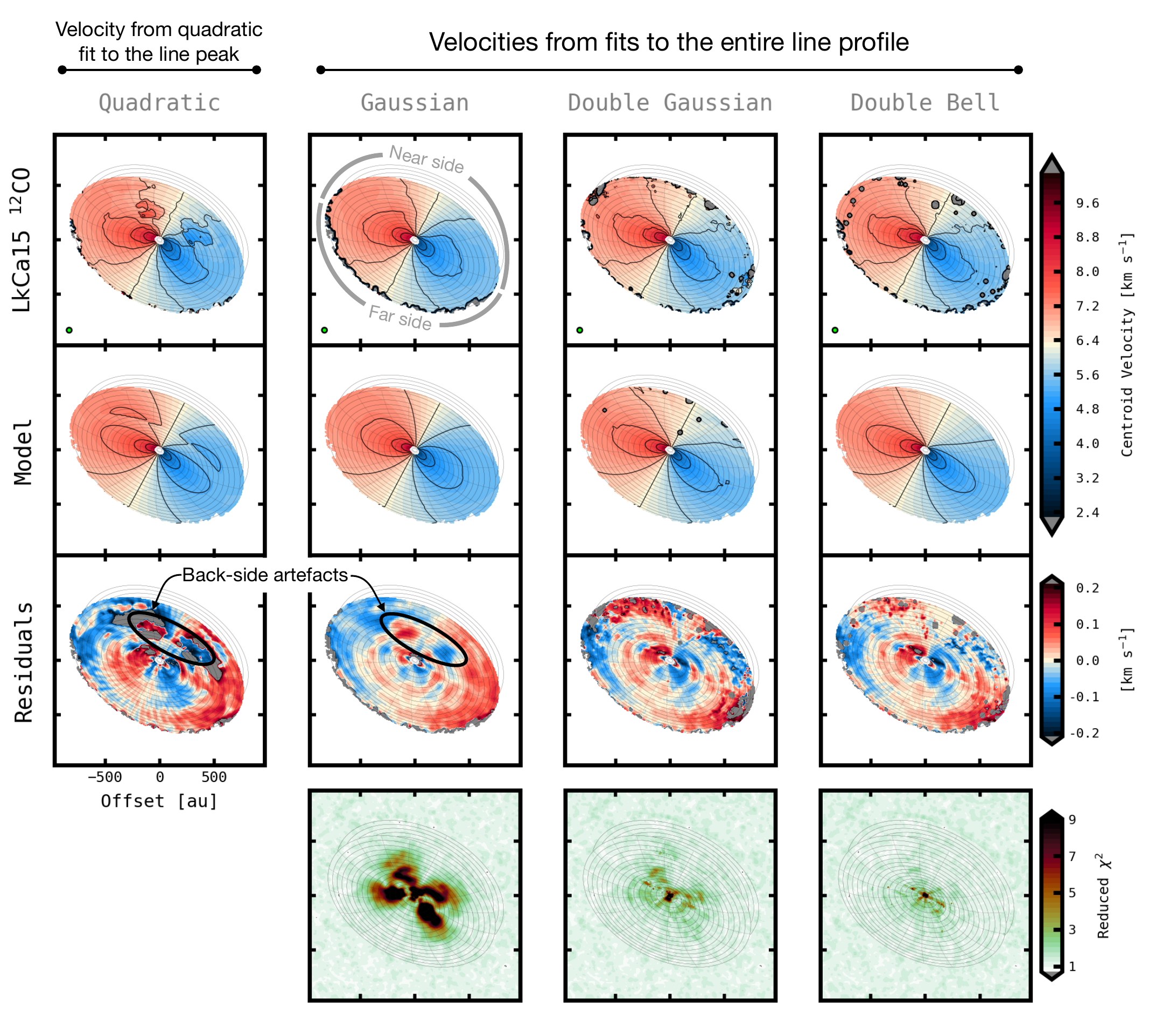}
      \caption{Comparison of velocity maps derived from single-- and double-component fits applied to the \twCO{} line profiles of the disc around \lkca{}. Quadratic fit velocity maps show significant backside contamination, particularly on the near side of the disc, and are susceptible to channelization effects as they consider a limited number of velocity points around the line peak for fitting, reducing their accuracy. Gaussian moments are accurate but exhibit reduced precision near the disc's diagonal axes due to backside contamination. Double-component moments are less affected by the back-side contribution, resulting in improved precision in the velocity maps and offering greater accuracy compared to the quadratic method, as they are less influenced by the channel spacing of the datacube. To evaluate the fit quality for the different kernels, the bottom row shows reduced $\chi^2$ values, calculated as $\chi^2_\nu = (n_{\rm ch} - n_{\rm pars})^{-1} \sum_{j}^{n_{\rm ch}} (I_{d,j} - I_{m,j})^2/s_j^2$ in the image plane,{where $s_j$ is a uniform weighting factor defined as the standard deviation of the observed intensity per channel, measured from line-free pixels.}
              }
         \label{fig:centroids_lkca15_12co}
   \end{figure*}

Fitting a two-component function to the disc emission can be challenging due to the wide range of velocities and intensities exhibited by these objects \citep[see also][]{casassus+2022, izquierdo+2022}. To address this, we use the best-fit model obtained with \discminer{} (introduced in Sect. \ref{sec:models}) to provide {the \textsc{curve\_fit} module with reliable} initial guesses for the peak intensity, line width, and centroid velocity of each emission surface in the disc on a per-pixel basis. This approach ensures that the line profile components are naturally assigned to either the front or back side of the disc based on their proximity in velocity and intensity to the corresponding emission surface properties derived from our smooth, Keplerian model of the source.

{Our algorithm incorporates an iterative routine in which the fit is reattempted for pixels where a single-component is identified, or for flagged pixels where: (a) neither a double-component nor a single-component fit succeeds, or (b) the fitted line width for either the front or back side is narrower than half the channel width. In each iteration, the fit uses the median parameter values from an $11\times11$ grid of neighboring pixels, where successful fits were obtained, as new initial guesses. We observe that the number of flagged pixels, which accounts for a small fraction ($<10$ per cent) of the total, stabilizes after 10 iterations. Figure \ref{fig:pixels_lkca15_12co} demonstrates this process for a region of the \lkca{} disc where the back side is warmer than the front side at the projected location in \twCO{} emission. Despite this, the algorithm successfully distinguishes between the two components and resolves the flagged pixels appropriately. } 

{Indeed, a key advantage of our model-informed, iterative approach is its capacity to mitigate confusion and prevent the emergence of unphysical substructures caused by projection effects in highly inclined sources, where the back side contribution to the line profile can often appear brighter than that of the front side in a substantial number of pixels. The ability to separate the front and back side components of the line profile significantly improves the precision and accuracy of the resulting velocity maps compared to those derived from methods relying on the line peak velocity or the centroid of Gaussian fits.}

To illustrate this, Figure \ref{fig:centroids_lkca15_12co} presents a comparison of velocity maps obtained from single-- and double-component fits applied to the \twCOfull{} line of the \lkca{} disc, along with the corresponding residuals relative to our \discminer{} model. As depicted in the figure, relying solely on a single-component fit, either quadratic or Gaussian, for this source, leads to contamination from the back side, especially on the near side of the disc where the lower emission surface contributes most prominently because of projection.

   \begin{figure*}
   \centering
    \includegraphics[width=1.0\textwidth]{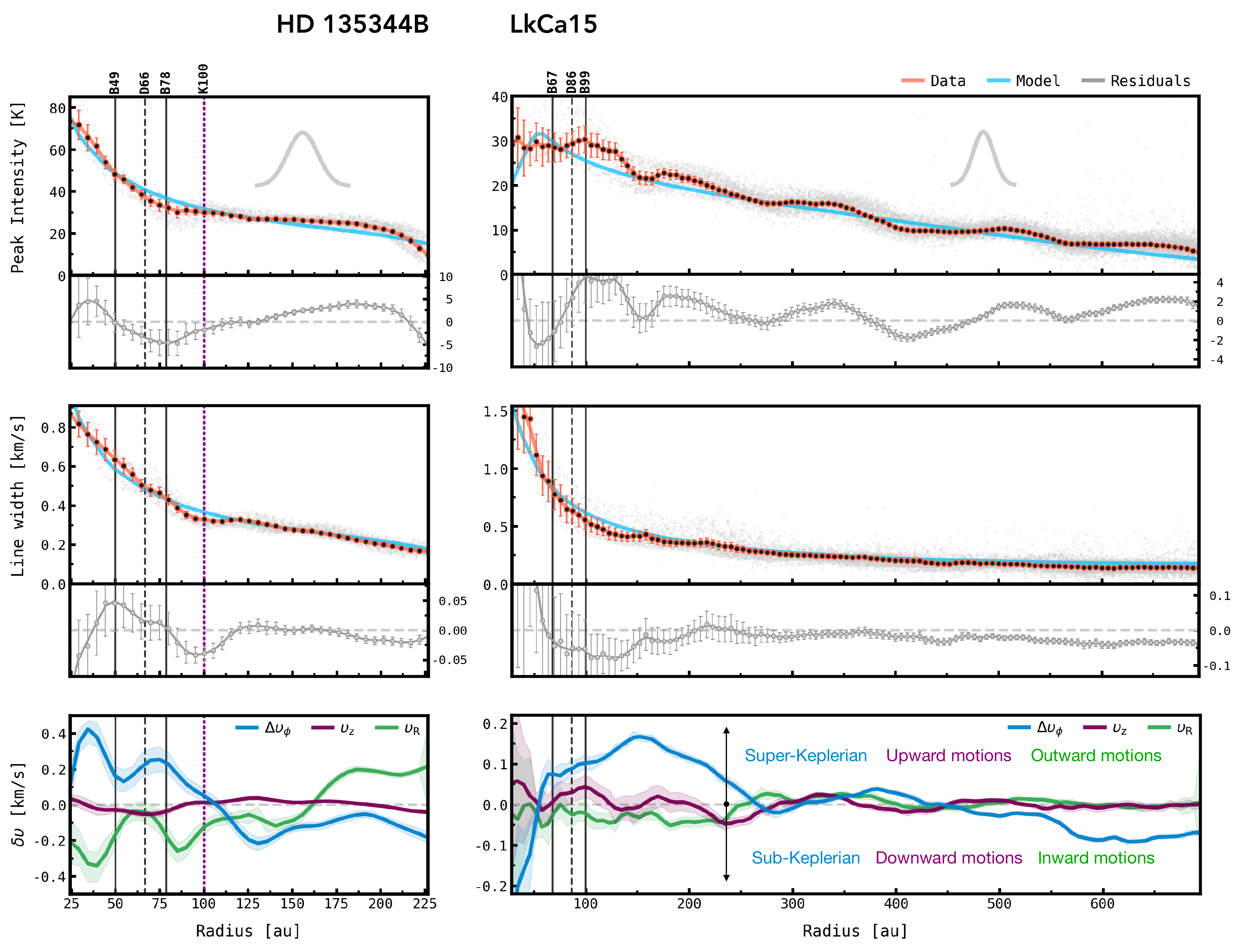}
      \caption{Peak intensity (top), line width (middle), and velocity profiles (bottom) extracted from \twCOfull{} emission for the front side of the discs of \hdone{} and \lkca{}, both imaged with a $0.''15$ beam, compared to the best-fit model quantities obtained with \discminer{}. The physical size of the beam's major axis is shown as a gray Gaussian profile in the top panels. Vertical lines indicate the radial locations of gaps (D) and rings (B) identified in the millimeter continuum by \citet{Curone_exoALMA}. Peak intensities have been converted to brightness temperature units via the Rayleigh-Jeans approximation. {Error bars in the top panels and shaded regions in the bottom panels represent the standard deviation divided by the square root of the number of independent beams along each projected annulus.} Gray points in the background of the peak intensity and line width profiles represent values extracted from all pixels in the image, mapped into the disc frame. 
              }
         \label{fig:radial_profiles_12co}
   \end{figure*}

\subsection{Velocity, intensity and linewidth profiles}

We calculate {radial profiles of peak intensity, line width, and three disc-frame components of velocity} by taking azimuthal averages on the respective moment maps derived from single-peak (Sect. \ref{sec:single_moments}) or double-peak fits (Sect. \ref{sec:double_moments}) applied to the observed molecular lines. 

{In cylindrical coordinates, the line-of-sight velocity, $\upsilon_0$, at a point $(R, \phi, z)$ in the disc can be expressed in its general form as,}
\begin{multline}
    \upsilon_{0}= \mathrm{sgn}_\mathrm{rot} \cdot \upsilon_\phi \cos{\phi} \sin{i}  - \upsilon_R \sin{\phi} \sin{i} \\ -\upsilon_z \cos{i} + \upsilon_{\rm LSRK}, \label{eq:v_los}
\end{multline}
{where the inclination $i$ can be negative or positive, $\mathrm{sgn}_\mathrm{rot}$ is positive for clockwise rotation (see Sect. \ref{sec:models} for details), the vertical component $\upsilon_z$ is positive for upward flows relative to the midplane, and the radial component $\upsilon_R$ is positive for outward motions relative to the disc centre.} 

{These velocity components, however, remain degenerate unless assumptions are made about the symmetry of the velocity field. To compute the rotation velocity component, $\upsilon_\phi$, at a given radial location, we employ the analytical method introduced by \citet{izquierdo+2023}, who demonstrated that if $\upsilon_{\phi}$ dominates over $\upsilon_z$ and $\upsilon_R$, it is proportional to the azimuthal average of the absolute value of the line-of-sight velocity map for an axisymmetric velocity field, and can be written as follows,}
\begin{equation} \label{eq:vphi}
    \upsilon_\phi = \frac{\psi}{4\sin{\frac{\psi}{4}}\sin{|i|}} \left<\left|\upsilon_{0} - \upsilon_{{\rm LSRK}}\right| \right>_{\psi},
\end{equation}
where $\psi$ denotes the angular extent of the azimuthal section over which the average is computed and {is required to be symmetric with respect to the disc's major axis.}

Also, the method states that the vertical velocity component, $\upsilon_{z}$, is proportional to the azimuthal average of the standard velocity map, 
\begin{equation}
    \upsilon_{z} = - \frac{1}{\cos{i}} \left<\upsilon_{0} - \upsilon_{{\rm LSRK}} \right>_{\psi}.
\end{equation}

Finally, {we introduce an estimate} of the radial velocity component, $\upsilon_{R}$, derived by subtracting the previously computed velocities, $\upsilon_{\phi}$ and $\upsilon_{z}$, projected along the line of sight, from the centroid velocity map, $\upsilon_{0}$, and taking the azimuthal average of the leftover residuals,
\begin{equation}
    \upsilon_{R} = -\left< A_\phi (\upsilon_0 - B_\phi \upsilon_{\phi} + \upsilon_z\cos{i} - \upsilon_{\rm LSRK})\right> _{\psi},
\end{equation}
where $A_\phi=(\sin{\phi} \sin{i})^{-1}$ and $B_\phi=\mathrm{sgn}_\mathrm{rot} \cos{\phi}\sin{i}$.

Employing the double-component moment maps introduced in Sect. \ref{sec:double_moments} for inclined sources allows us to use the entire azimuthal extent of the disc when calculating these averages, thereby reducing the impact of small-scale fluctuations on the rotation, vertical, and radial velocity profiles and enhancing the accuracy of our measurements. In contrast, single-component fits of these sources often require extensive spatial masking to eliminate back-side contamination. 

We highlight that the secondary component of the line profiles derived from the double-bell fit (indicated by the hatched magenta profiles in Fig. \ref{fig:line_profiles_lkca15_12co}) is also helpful as it can be used to extract physical information from the back side of the disc, which is closely linked to the vertical snowline of the molecular tracer \citep[see e.g.,][]{pinte+2018_surfaces}. This component provides well-constrained peak intensity profiles, which are proxies of the back side temperature structure, as well as rotation velocities. 
However, caution is necessary with the latter as our models do not account for absorption from the midplane \citep[see e.g.][]{pinte+2018_surfaces}, which may induce fluctuations in the retrieved back-side line widths and centroids that are difficult to predict.

Figure \ref{fig:radial_profiles_12co} illustrates the peak intensity, line width, and {velocity profiles} extracted from \twCO{} emission for the front side of the discs of \hdone{} and \lkca{}.
The peak intensity profiles of both discs show a monotonically decreasing trend, with local variations on the order of a few Kelvin and a typical scale size of $\sim\!50$\,au in radius. 
A detailed analysis of the three-dimensional temperature structure across all \exoalma{} discs is presented in \citet{Galloway_exoALMA}. Additionally, \citet{Rosotti_exoALMA} estimate the surface densities and masses of these discs by examining their temperatures and CO emission surface heights.  

We also note that both discs exhibit rotation velocities that are close to Keplerian, but not perfectly so. Analyses of radial deviations from Keplerian motion across all \exoalma{} targets are presented in \citet{Stadler_exoALMA} and \citet{Longarini_exoALMA}, where these variations are linked to modulations in the pressure structure and the self-gravity of the gas discs. 

\section{Summary} \label{sec:conclusions}

In this paper, we present a methodology to investigate the radial gas structure and substructure in the 15 protoplanetary discs targeted by the ALMA large program \exoalma{}. Our approach is based on the \discminer{} modelling framework and incorporates an improved iterative two-component fitting routine devised for inclined sources, allowing us to separate the contribution of the discs' front and back sides to the total intensity profile, which is crucial for a precise characterization of the dynamical and physical structure of our sources. We report best-fit parameters describing the orientation and vertical structure of our targets, along with Keplerian stellar masses, which serve as a foundation for subsequent analysis papers in the \exoalma{} series. Our study underscores the critical role of high-quality data and meticulous examination of molecular line profiles in advancing our understanding of protoplanetary disc physics and dynamics. 

\section*{Acknowledgments}

We thank the anonymous referee for their comments and suggestions which greatly improved the quality of this work.
This paper makes use of the following ALMA data: ADS/JAO.ALMA\#2021.1.01123.L. ALMA is a partnership of ESO (representing its member states), NSF (USA) and NINS (Japan), together with NRC (Canada), MOST and ASIAA (Taiwan), and KASI (Republic of Korea), in cooperation with the Republic of Chile. The Joint ALMA Observatory is operated by ESO, AUI/NRAO and NAOJ. The National Radio Astronomy Observatory is a facility of the National Science Foundation operated under cooperative agreement by Associated Universities, Inc. We thank the North American ALMA Science Center (NAASC) for their generous support including providing computing facilities and financial support for student attendance at workshops and publications. Support for AFI was provided by NASA through the NASA Hubble Fellowship grant No. HST-HF2-51532.001-A awarded by the Space Telescope Science Institute, which is operated by the Association of Universities for Research in Astronomy, Inc., for NASA, under contract NAS5-26555. JS, MB, DF have received funding from the European Research Council (ERC) under the European Union’s Horizon 2020 research and innovation programme (PROTOPLANETS, grant agreement No. 101002188). Computations by JS have been performed on the `Mesocentre SIGAMM' machine, hosted by Observatoire de la Cote d’Azur. CP acknowledges Australian Research Council funding via FT170100040, DP18010423, DP220103767, and DP240103290. JB acknowledges support from NASA XRP grant No. 80NSSC23K1312. SF is funded by the European Union (ERC, UNVEIL, 101076613), and acknowledges financial contribution from PRIN-MUR 2022YP5ACE. CL has received funding from the European Union's Horizon 2020 research and innovation program under the Marie Sklodowska-Curie grant agreement No. 823823 (DUSTBUSTERS) and by the UK Science and Technology research Council (STFC) via the consolidated grant ST/W000997/1. PC and LT acknowledge support by the Italian Ministero dell'Istruzione, Universit\`a e Ricerca through the grant Progetti Premiali 2012 – iALMA (CUP C52I13000140001) and by the ANID BASAL project FB210003. NC has received funding from the European Research Council (ERC) under the European Union Horizon Europe research and innovation program (grant agreement No. 101042275, project Stellar-MADE). M. Flock has received funding from the European Research Council (ERC) under the European Unions Horizon 2020 research and innovation program (grant agreement No. 757957). M. Fukagawa is supported by a Grant-in-Aid from the Japan Society for the Promotion of Science (KAKENHI: No. JP22H01274). CH acknowledges support from NSF AAG grant No. 2407679. IH and TH are supported by an Australian Government Research Training Program (RTP) Scholarship. JDI acknowledges support from an STFC Ernest Rutherford Fellowship (ST/W004119/1) and a University Academic Fellowship from the University of Leeds. AI acknowledges support from the National Aeronautics and Space Administration under grant No. 80NSSC18K0828.
G. Lodato has received funding from the European Union's Horizon 2020 research and innovation program under the Marie Sklodowska-Curie grant agreement No. 823823 (DUSTBUSTERS). DP acknowledges Australian Research Council funding via DP18010423, DP220103767, and DP240103290. GR acknowledges funding from the Fondazione Cariplo, grant no. 2022-1217, and the European Research Council (ERC) under the European Union’s Horizon Europe Research \& Innovation Programme under grant agreement no. 101039651 (DiscEvol). H-WY acknowledges support from National Science and Technology Council (NSTC) in Taiwan through grant NSTC 113-2112-M-001-035- and from the Academia Sinica Career Development Award (AS-CDA-111-M03). GWF acknowledges support from the European Research Council (ERC) under the European Union Horizon 2020 research and innovation program (Grant agreement no. 815559 (MHDiscs)). GWF was granted access to the HPC resources of IDRIS under the allocation A0120402231 made by GENCI. AJW has received funding from the European Union's Horizon 2020
research and innovation programme under the Marie Skłodowska-Curie grant
agreement No 101104656. TCY acknowledges support by Grant-in-Aid for JSPS Fellows JP23KJ1008. Support for BZ was provided by The Brinson Foundation. This work was partly supported by the Deutsche Forschungsgemein- schaft (DFG, German Research Foundation) - Ref no. 325594231 FOR 2634/2 TE 1024/2-1, and by the DFG Cluster of Excellence Origins (www.origins-cluster.de). This project has received funding from the European Research Council (ERC) via the ERC Synergy Grant ECOGAL (grant 855130). Views and opinions expressed by ERC-funded scientists are however those of the author(s) only and do not necessarily reflect those of the European Union or the European Research Council. Neither the European Union nor the granting authority can be held responsible for them.  

\software{\textsc{astropy} \citep{Astropy_2022}, \textsc{casa} \citep{casa}, \textsc{cmasher} \citep{cmasher+2020}, \discminer{} \citep{izquierdo+2021}, \textsc{emcee} \citep{foreman+2013}, \textsc{matplotlib} \citep{Hunter_mpl}, \textsc{numpy} \citep{harris_np}, \textsc{scikit-image} \citep{scikit-image}, \textsc{scipy} \citep{Virtanen_scipy}, \textsc{radio-beam} \citep{radio-beam}, \textsc{spectral-cube} \citep{spectral-cube}}

\bibliography{main}{}
\bibliographystyle{aasjournal}

\appendix

\section{Supporting figures} \label{sec:supporting_figures}

\begin{figure*}
   \centering
    \includegraphics[width=0.8\textwidth]{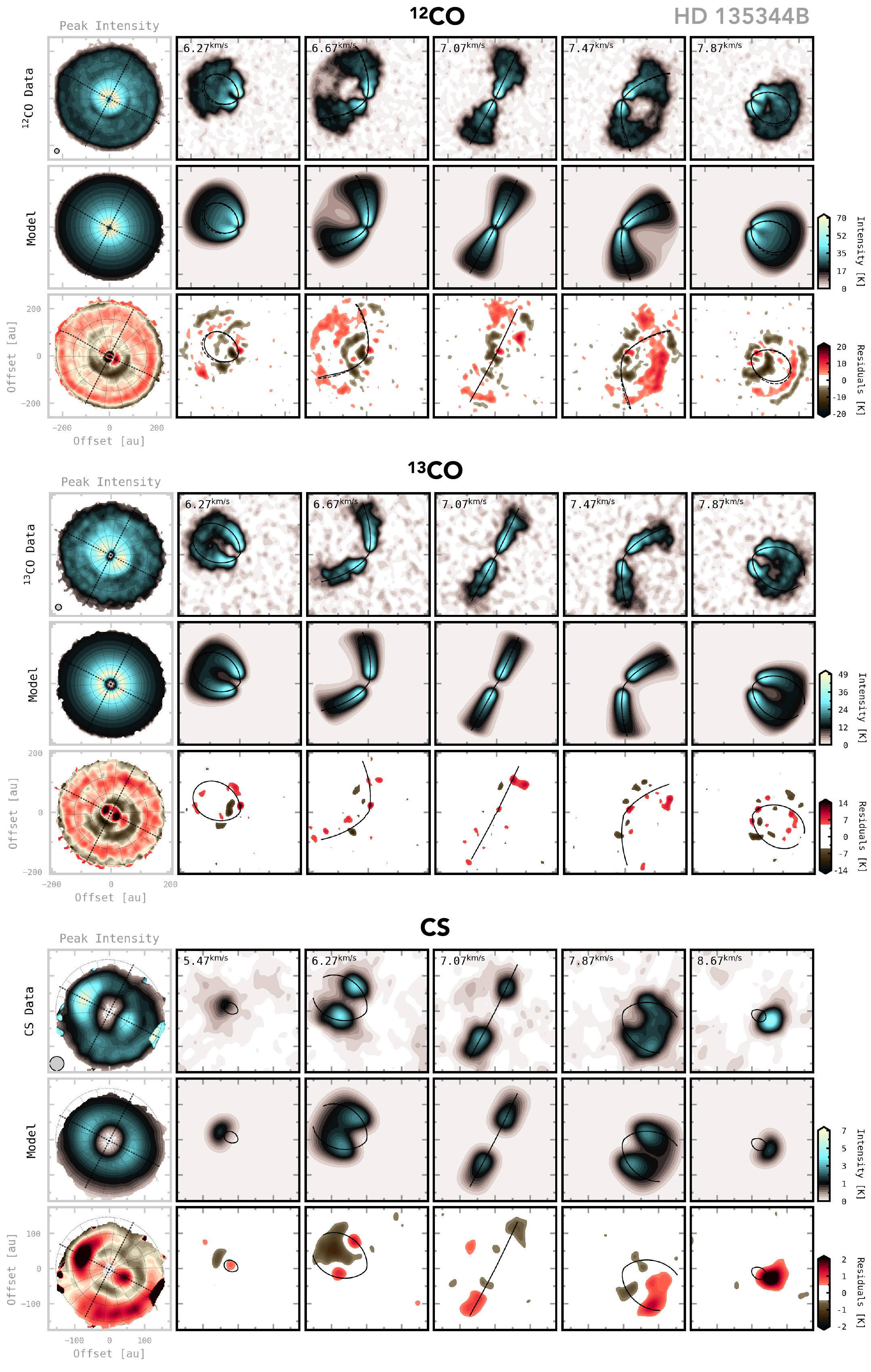}
      \caption{Selected \twCO{} (top), \thCO{} (middle), and CS (bottom) intensity channels for the disc of \hdone{}. Each panel displays, from top to bottom, the channel maps for the data, the best-fit model, and the corresponding residuals, with residuals below three times the rms noise whited out. Peak intensity maps and residuals are shown in the leftmost column of each panel. Solid and dashed lines overlaid on the channel maps represent isovelocity contours from the model's front and back sides. 
              }
         \label{fig:channel_maps_hd135344}
   \end{figure*}

\begin{figure*}
   \centering
    \includegraphics[width=0.8\textwidth]{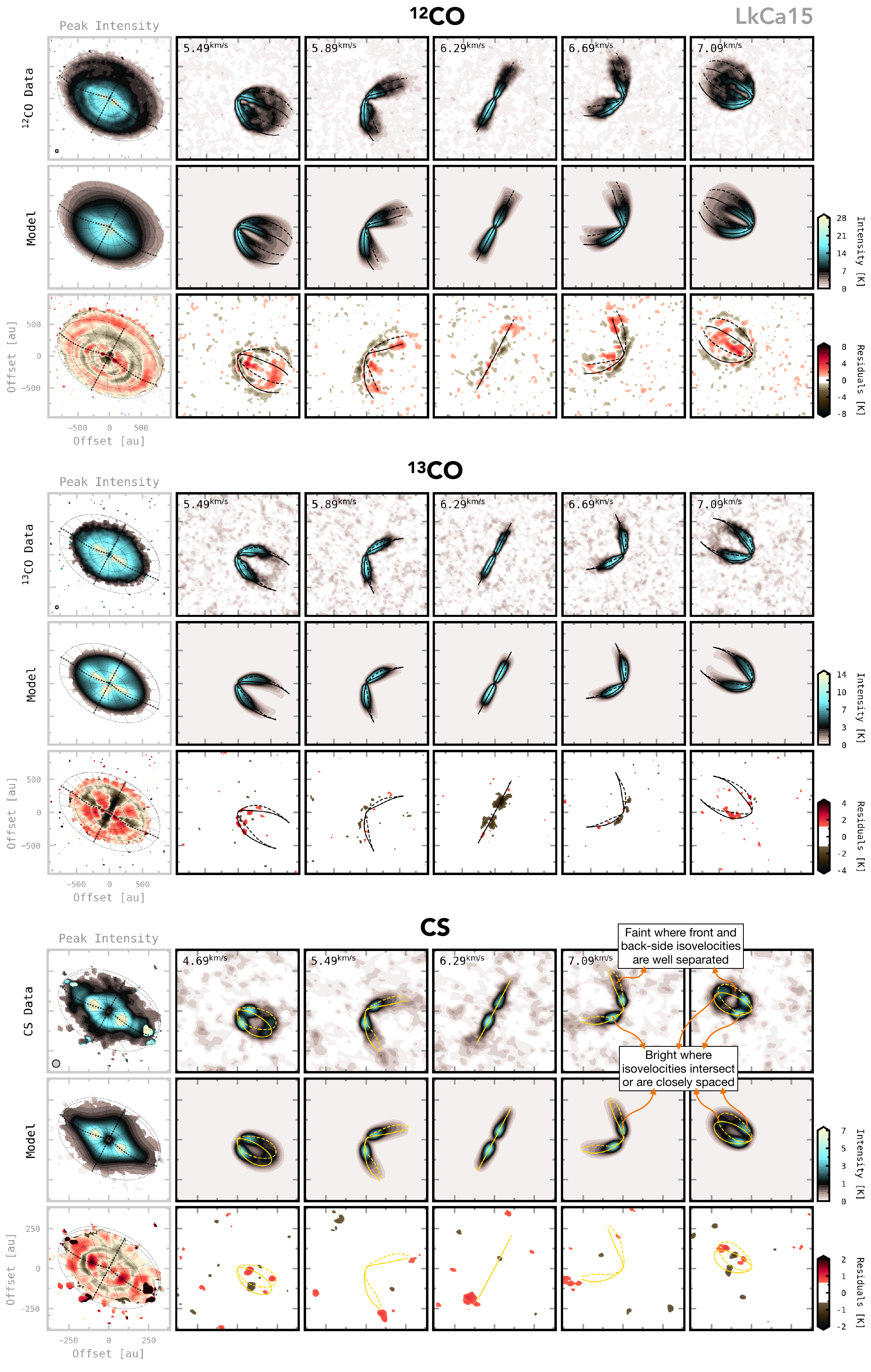}
      \caption{Same as Figure \ref{fig:channel_maps_hd135344}, but for the disc of \lkca{}. The \thCO{} and CS models are assumed to be optically thin and therefore use a `sum' kernel to combine the contributions of the front and back sides to the total intensity (see Sects. \ref{sec:models} and \ref{sec:single_moments} for details). CS isovelocity contours are shown in yellow to highlight regions where increased intensity results from overlapping column densities of the front and back sides at matching line-of-sight velocities.
              }
         \label{fig:channel_maps_lkca15}
   \end{figure*}

\begin{figure*}
   \centering
    \includegraphics[width=0.9\textwidth]{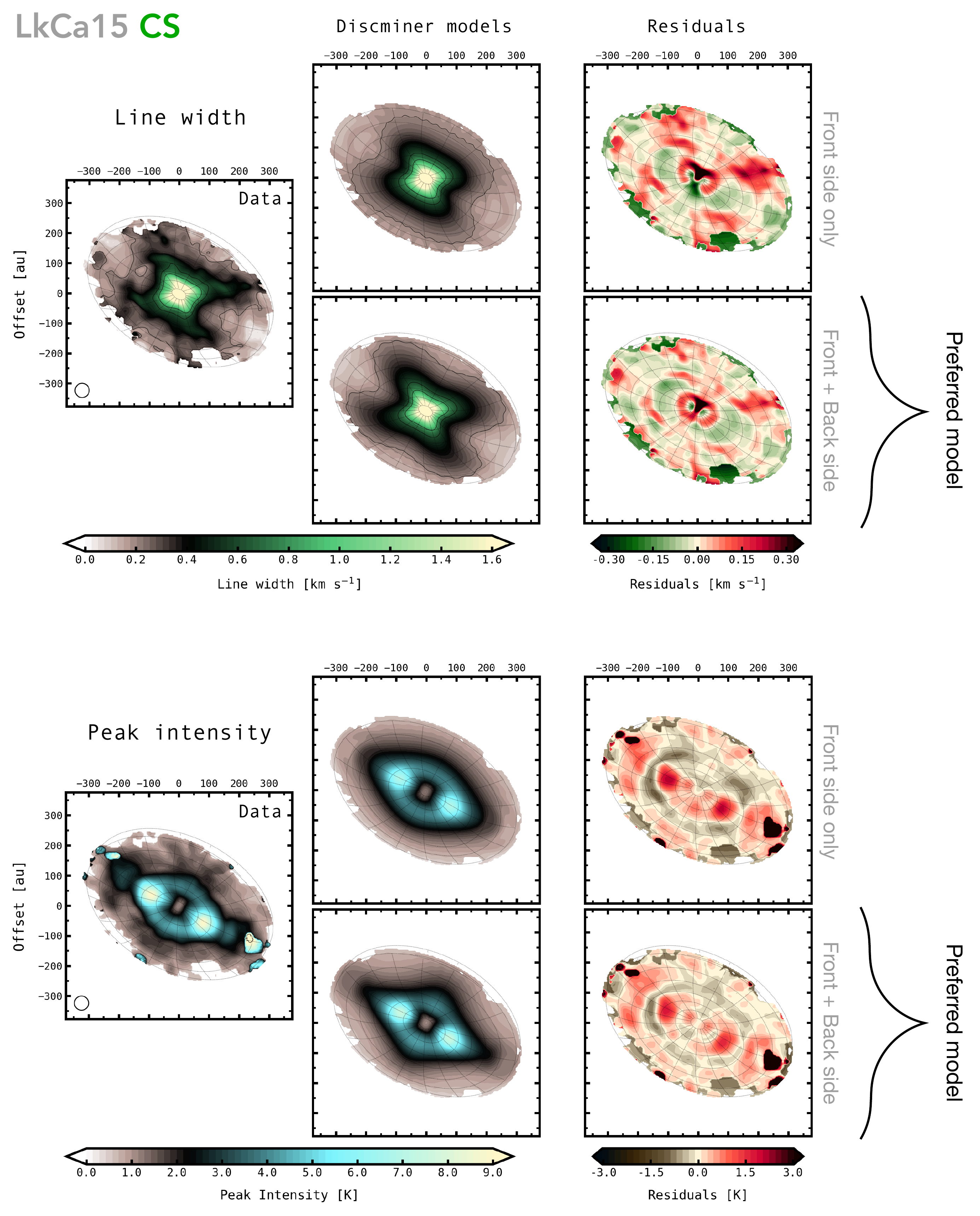}
      \caption{Comparison of line width and peak intensity residuals obtained from \CSfull{} emission for the disc of \lkca{} using two models. The first model (top rows) considers only the contribution of the upper surface to the total intensity profile, while the second model (bottom rows) incorporates the summed contribution from both upper and lower emission surfaces. The latter model  provides a better reproduction of line width signatures along the disc's diagonal axes, and intensity features along the disc's main axes, which result in lower line width and intensity residuals as illustrated in the rightmost panels.
              }
         \label{fig:linewidth_peakint_lkca15_cs}
   \end{figure*}

In this Appendix, we provide additional figures to support the discussions in Sections \ref{sec:models} and \ref{sec:observables}. Figures \ref{fig:channel_maps_hd135344} and \ref{fig:channel_maps_lkca15} show selected intensity channels of \twCO{}, \thCO{}, and CS for \hdone{} and \lkca{}, along with the corresponding best-fit models and residuals. Figure \ref{fig:linewidth_peakint_lkca15_cs} compares intensity and line width residuals in CS emission for the \lkca{} disc, contrasting frontside-only and front+backside models, and motivating our choice of the latter.

\end{document}